\documentclass[
aps,
amsmath,
amssymb,
preprint,
author-year,
]{revtex4-1}

\usepackage{dcolumn}
\usepackage{bm}
\usepackage{amsmath,amssymb,fancybox,amscd,ascmac,accents}
\usepackage[all]{xy}
\usepackage{braket}
\usepackage{makeidx}
\usepackage{mathrsfs}
\usepackage{graphicx}

\newcommand{\dis}{\displaystyle}

\newcommand{\pa}{\partial}
\newcommand{\na}{\nabla}
\newcommand{\ga}{\gamma}
\newcommand{\Ga}{\Gamma}
\newcommand{\lam}{\lambda}

\newcommand{\sig}{\sigma}

\newcommand{\lr}{\leftrightarrow}

\newcommand{\no}{\nonumber}

\def\bee{\begin{equation}}
\def\ede{\end{equation}}
\def\bna{\begin{eqnarray}}
\def\ena{\end{eqnarray}}
\def\bgen{\begin{enumerate}}
\def\eden{\end{enumerate}}
\def\bede{\begin{description}}
\def\ende{\end{description}}
\def\bex{\begin{axio}}
\def\enx{\end{axio}}
\def\bef{\begin{defi}}
\def\enf{\end{defi}}
\def\bepp{\begin{prop}}
\def\enpp{\end{prop}}
\def\bet{\begin{theo}}
\def\ent{\end{theo}}
\def\bel{\begin{lemm}}
\def\enl{\end{lemm}}
\def\bep{\begin{proo}}
\def\enp{\end{proo}}
\def\bepb{\begin{prob}}
\def\enpb{\end{prob}}
\def\bpri{\begin{prin}}
\def\epri{\end{prin}}

\begin{document}

\preprint{KOBE-TH-15-14}
\title{Quasi-Normal Modes of Black Holes in Lovelock Gravity}

\author{Daiske Yoshida}
 \email{154s129s@stu.kobe-u.ac.jp}
 \affiliation{Department of Physics, Kobe University, Kobe, 657-8501, Japan}
 
\author{Jiro Soda}
 \email{jiro@phys.sci.kobe-u.ac.jp}
 \affiliation{Department of Physics, Kobe University, Kobe, 657-8501, Japan}

\date{\today}

\begin{abstract}
We study quasinormal modes of black holes in Lovelock gravity. 
 We formulate the WKB method adapted to Lovelock gravity for the calculation of quasinormal frequencies (QNFs). 
 As a demonstration, we calculate various QNFs of Lovelock black holes in seven and eight dimensions. 
 We find that the QNFs show remarkable features depending on the coefficients of the Lovelock terms,
 the species of perturbations, and spacetime dimensions. 
 In the case of the scalar field, when we increase the coefficient of the third order Lovelock term, the real part of QNFs increases, but the decay rate becomes small irrespective of the mass of the black hole.
 For small black holes, the decay rate ceases to depend on the Gauss-Bonnet term.   
 In the case of tensor type perturbations of the metric field, the tendency of the real part of QNFs is opposite to that of the scalar field.
 The QNFs of vector type perturbations of the metric show no particular behavior.
 The behavior of QNFs of the scalar type perturbations of the metric field is similar to the vector type. 
 However, available data are  rather sparse, which indicates that the WKB method is not applicable to many models for this sector.
\end{abstract}

%Use showkeys class option if keyword
\maketitle

%\begin{quotation}
%The ``lead paragraph'' is encapsulated with the \LaTeX\ 
%\verb+quotation+ environment and is formatted as a single paragraph before the first section heading. 
%(The \verb+quotation+ environment reverts to its usual meaning after the first sectioning command.) 
%Note that numbered references are allowed in the lead paragraph.
%The lead paragraph will only be found in an article being prepared for the journal \textit{Chaos}.
%\end{quotation}

\section{Introduction}
According to the large extra-dimension scenario, there exists a chance for higher dimensional black holes to be created at the LHC~\cite{Giddings:2001bu,Giddings:2008gr}.
 Hence, the higher dimensional black holes have been intensively investigated.
 It should be noted that the Einstein theory of gravity is not the most general one in higher dimensions.  
 In four dimensions, the Einstein gravity can be deduced by assuming the general coordinate covariance and the absence of higher derivative terms larger than the second order in the Lagrangian.
 In higher dimensions, however, the same assumptions lead to Lovelock theory of gravity~\cite{Lovelock:1971yv}.
 In addition to this mathematical motivation, we have a physical motivation to consider Lovelock gravity. 
 In fact, at the energy scale of black hole production, the Einstein gravity is not reliable any more. 
 It is widely believed that string theory is valid at the fundamental scale.
 Remarkably, string theory can be consistently formulated only in ten dimensions.
 As is well known,  string theory reduces to Einstein gravity in the low energy limit. 
 In string theory, however, there are higher curvature corrections in addition to the Einstein-Hilbert term.
 Thus, it is natural to extend Einstein gravity into those with higher power of curvature in higher dimensions.
 It is Lovelock gravity that belongs to such class of theories. 
 In Lovelock gravity, it is known that there exist static spherically symmetric black holes~\cite{Boulware:1985wk,Charmousis:2008kc,Wheeler:1985qd}.
 Hence, it is reasonable to suppose black holes produced at the LHC are of this type.

In five or six dimensions, the Lovelock gravity reduces to the so-called Gauss-Bonnet gravity which has static and spherically symmetric
 black hole solutions~\cite{Boulware:1985wk,Charmousis:2008kc,Wheeler:1985qd}. 
 The stability analysis of Gauss-Bonnet black holes under scalar, vector, and tensor perturbations has been performed~\cite{Dotti:2004sh,Dotti:2005sq, Gleiser:2005ra}. 
 It is shown that there exists the scalar mode instability in five dimensions, the tensor mode instability in six dimensions, and no instability in other dimensions~\cite{Beroiz:2007gp}. 
 In more than six dimensions, however, we need to consider more general Lovelock gravity. 
 For example, when we consider ten dimensional black holes, we need to incorporate the third and fourth order Lovelock terms into the action.
 Indeed, when we consider black holes at the LHC, it is important to consider these higher order Lovelock terms. 
 The stability of black holes in any order Lovelock gravity has been studied~\cite{Takahashi:2009dz,Takahashi:2009xh,Takahashi:2010ye,Takahashi:2010gz,Takahashi:2011du}.
 It turned out that small black holes are unstable in any dimensions.

In order to understand properties of black holes in Lovelock gravity, it is important to study QNMs of black holes. 
 The QNFs of Gauss-Bonnet gravity have already been investigated~\cite{Konoplya:2008ix,Chen:2015fuf,Chen:2009an}.
 Thus, the aim of this paper is to calculate QNFs of the stable large black holes in Lovelock gravity using the WKB method~\cite{Schutz:1985zz,Konoplya:2011qq}.
 The QNFs of Lovelock black holes have been calculated for the special background~\cite{Prasobh:2013nda,Prasobh:2014zea,Lin:2014vla}.
 Since the metric is analytically known for such cases, there is no difficulty in using the  WKB-method for obtaining QNFs.
 In general, however, a problem arises because the metric of the black hole is only implicitly given by an algebraic equation. 
 Hence, the primary aim of this paper is to give a general formalism to calculate QNFs of black holes in Lovelock gravity.
 The other purpose of this paper is to calculate QNFs of Lovelock black holes in seven and eight dimensions and reveal effects of higher order Lovelock terms on QNFs.

The organization of the paper is as follows. In Sec. II, we introduce Lovelock gravity and Lovelock black holes.
 In Sec. III, we review the perturbation theory of Lovelock black holes. 
 In Sec. IV, we explain the  WKB method for obtaining QNFs. 
 There, we propose a novel method adapted to Lovelock gravity for calculating QNFs.
 Then, we present numerical results  and extract many interesting features from the results.
 The final section is devoted to the conclusion.

\section{\label{sec:level2} Black holes in Lovelock Gravity}

In this section, we briefly review Lovelock gravity in higher dimensions~\cite{Lovelock:1971yv} and derive static black hole solutions determined by an algebraic equation. 

Lovelock gravity is characterized by the general coordinate covariance and the absence of terms with higher than the second order derivatives in the equations of motion.
 The Lagrangian satisfying these properties can be constructed  from $m$th Lovelock terms $\mathcal{L}_{(m)}$ defined by the product of Riemann tensors
\bee
\mathcal{L}_{(m)} \equiv \delta^{\lam_1 \sig_1 \cdots \lam_m \sig_m}_{\rho_1 \kappa_1 \cdots \rho_m \kappa_m} R^{\rho_1 \kappa_1}_{\hspace{4mm} \lam_1 \sig_1} \cdots R^{\rho_m \kappa_m}_{\hspace{4mm} \lam_m \sig_m} \ ,
\ede
where we used the totally antisymmetric Kronecker delta
\bee
\delta^{\lam_{1} \cdots \lam_{p}}_{\rho_1 \cdots \rho_{p}} \equiv \left| \begin{array}{cccc} \delta^{\lam_1}_{\kappa_1} & \delta^{\lam_2}_{\kappa_1} & \cdots & \delta^{\lam_p}_{\kappa_1} \\
\delta^{\lam_1}_{\kappa_2} & \delta^{\lam_2}_{\kappa_2} & \cdots & \delta^{\lam_p}_{\kappa_2} \\  \vdots & \vdots & \ddots & \vdots \\  
\delta^{\lam_1}_{\kappa_p} & \delta^{\lam_2}_{\kappa_p} & \cdots & \delta^{\lam_p}_{\kappa_p} \end{array}\right|  \ .
\ede
In $D$-dimensions, Lagrangian density $\mathcal{L}_D$ is written by
\bee
\mathcal{L}_D \equiv -2 \Lambda + \sum^{k}_{m=1} \frac{a_m}{m \cdot 2^{m+1}} \mathcal{L}_{(m)} \ ,
\ede
where $\Lambda$ is a cosmological constant, $a_m$ represents the coupling constants of Lovelock gravity and $k$ is a parameter defined by
\bee
k \equiv \left[ \frac{D-1}{2} \right] .
\ede
 This Lagrangian is called the Lovelock Lagrangian.
 We can set $a_{1} = 1$ without losing generality.
 The action obviously has the general coordinate invariance.
 It is also straightforward to see the absence of higher derivative terms larger than the second order derivatives.
 Performing the  variation with respect to the metric, we obtain the Lovelock tensor $\mathcal{G}^{D \,\,\nu}_{\mu}$  defined as
\bee
\mathcal{G}^{D \,\,\nu}_{\mu} = \Lambda \delta^{\nu}_{\mu} - \sum^k_{m=1} \frac{a_m}{m \cdot 2^{m+1}} \delta^{\nu \lam_1 \sig_1 \cdots \lam_m \sig_m}_{\mu \rho_1 \kappa_1 \cdots \rho_m \kappa_m} R^{\rho_1 \kappa_1}_{\hspace{4mm} \lam_1 \sig_1} \cdots R^{\rho_m \kappa_m}_{\hspace{4mm} \lam_m \sig_m} \ ,
\ede
where we used the Bianchi indentity to eliminate the terms with derivative of the Riemann tensor.
 This tensor is equal to the Einstein tensor in $D=4$ and the Einstein-Gauss-Bonnet tensor in $D=5,6$.
 Thus, Lovelock theory can be regarded as a natural generalization of  Einstein theory.

It is known that there exist static black hole solutions in Lovelock gravity~\cite{Boulware:1985wk,Wheeler:1985qd,Charmousis:2008kc}. To obtain the solutions, we put an ansatz
\bee
ds^2 = g_{\mu\nu} \mathrm{d}x^{\mu} \mathrm{d}x^{\nu} =- f(r) \mathrm{d}t^2 
+ \frac{1}{f(r)} \mathrm{d}r^2 + r^2 \ga_{ij} \, \mathrm{d}x^i \mathrm{d}x^j \ ,
\ede
where  $\ga_{ij}$ is the metric of the sphere in $n \equiv D-2$ dimensions.
 Using this ansatz, we can  calculate the Lovelock tensor which must be zero in vacuum.
 It is convenient to define a new variable $\psi(r)$ as
\bee
\psi(r) = \frac{1-f(r)}{r^2} \ .
\ede
 Then, we can integrate equations of motion and obtain 
\bee
\mathcal{P}(\psi) = \frac{\mathcal{M}}{r^{n+1}} \ ,
\label{algebraic_eq}
\ede 
where the polynomial $\mathcal{P}(\psi)$ is defined as 
\bee
\mathcal{P}(\psi) = \left\{ \begin{array}{c} \dis \psi - \frac{2 \Lambda}{n(n+1)} \hspace{4.9cm} (k=1) \ , \\ 
                            \\
                            \dis \sum^k_{q=2} \left[ \frac{a_q \prod^{2q-2}_{p=1}(n-p)}{q} \psi^q \right] + \psi - \frac{2 \Lambda}{n(n+1)} \hspace{5mm} (k \geq 2) \ .
                            \end{array} \right. \label{kettei}  
\ede
 Here, $\mathcal{M}$ is related to ADM mass of black hole as follows:  
\bee
\mathcal{M} \equiv \frac{\Ga (\frac{n+1}{2})}{2 \pi^{\frac{n+1}{2}}} M_{ADM} \ .
\ede

In this paper, we will focus on the asymptotically flat spacetime.
 To realize the asymptotically flat spacetime, we set $\Lambda = 0$ and $a_m > 0$ for all $m$. 
 Thus, $\mathcal{P}(\psi)$ becomes a monotonically increasing function of $\psi$ with zero at $\psi=0$. 
 Since $\psi=0$ is achieved at $r=\infty$, we can realize asymptotically flat spacetime with this setup. 
 It should be stressed that Eq.(\ref{algebraic_eq}) implicitly define $\psi (r)$, namely, the metric function $f(r)$.
 In other words, we cannot obtain the metric in an analytic form in general.

\section{Master Equations}
In the previous section, we have presented black hole solutions in Lovelock gravity. Now, we consider a scalar field on this spacetime and linear metric perturbations around this background spacetime. 
 We show that there exists a master equation in any case.

The master equation is an equation for the master variable by which all other components of the metric can be derived.
 In the case of the scalar field, the master equation is nothing but the Klein-Gordon equation. 
 In the case of metric perturbations, one can also obtain the master equations.
 The explicit relation between the master variable and the other metric components can be found  in \cite{Takahashi:2010ye}.

Once we obtain the master equations which are separable, we can analyze the stability of black holes and calculate QNFs of black holes.
 Since we are considering static background, the time-dependence of the fields can be separated by the factor
\bee
e^{-i \omega t} \ .
\ede
 Moreover, due to the spherical symmetry, angular dependence of the master variable can be  separated by using harmonics on the sphere.
 Thus, the master equation can be reduced to  the  Schr$\ddot{o}$dinger type equation for the radial direction as
\bee
\frac{d^2 \Phi(x)}{dx^2} + \left[ \omega^{2} - V(x) \right]\Phi(x) = 0 \ ,
\ede
where we have defined the tortoise coordinate $x$ as follows:
\bee
\frac{\mathrm{d}x}{\mathrm{d}r} = \frac{1}{f(r)} \ ,
\ede

For quasinormal mode analysis, the shape of the effective potential $V(x)$ is our main objective.
 In the rest of this section, we write down the various effective potentials we will use in this paper. 

\subsection{Effective potential of scalar field}
As we mentioned, the master equation of a scalar field is the Klein-Gordon equation. 
 The Klein-Gordon equation with the mass $\mu$ in curved spacetime reads
\bee
\left[ \na_{\mu} \na^{\mu} - \mu^{2} \right]\phi(t,r, \Omega) = 0 \ ,
\ede
where $\Omega$ represents the coordinates in a $n$-dimensional sphere.
 The spectrum of spherical harmonics $Y_L$ on a $n$-dimensional sphere is given by
\bee
\Delta Y_L (\Omega ) =  - L(L+n-1) Y_L (\Omega ) \ ,
\ede
with an integer $L\geq 0$.
 After separating variables, we obtain the following  equation 
\bee
\frac{d^2 \Phi(x)}{dx^2} + \left[ \omega^{2} - f(r) \left( \mu^{2} + f(r) \frac{n(n-2)}{4 r^{2}} + f'(r) \frac{n}{2 r} + \frac{L(L+n-1)}{r^{2}} \right) \right] \Phi(x) = 0 \ .
\ede
 Here, a prime represents a differentiation with respect to the radial coordinate $r$.
 Therefore, in this case we get the effective potential as follows:
\bee
V(r) = f(r) \left( \mu^{2} + f(r) \frac{n(n-2)}{4 r^{2}} + f'(r) \frac{n}{2 r} + \frac{L(L+n-1)}{r^{2}} \right) \ .
\ede

\subsection{Effective potential of metric perturbations}
The metric perturbations in Lovelock gravity was fully analyzed in \cite{Takahashi:2010ye}. There, the metric perturbations are classified into tensor, vector, and scalar types according to the symmetry on the sphere.
 Then, the master equations for metric perturbations of tensor-, vector-, and scalar-type have been derived.
 In this article, we simply use the effective potentials obtained in \cite{Takahashi:2010ye}. 
 In the following, we use the functions $T(r)$ and $X(r)$  defined by
\bna
T(r) &\equiv& r^{n-1} \pa_{\psi} \mathcal{P}(\psi) \ , \\
X(r) &\equiv& \frac{-2nf(r) + 2 L(L+n-1) + n r f'(r) }{r \sqrt{T'(r)}} \ .
\ena
 For the absence of the ghost, we need to assume $T'(r)>0$ for $^{\forall}r>0$. 
 We call it the $T'$-condition. 

\subsubsection{Tensor-type perturbations}
The master equation for tensor type perturbations can be written as
\bee
V_{t}(r) = f(r) \left[ \frac{1}{r \sqrt{T'(r)}}  \left( f(r) \left( r \sqrt{T'(r)} \right)' \right)' + \frac{ L(L+n-1) }{ (n-2) }\frac{1}{r} \left( \ln{ T'(r) } \right)' \right]  \ .
\ede
 It is known that there exists the instability of small black holes under tensor perturbations in even dimensions.

\subsubsection{Vector-type perturbations}

The master equation for vector perturbations reads
\bee
V_{v}(r) = f(r) \left[ r \sqrt{T'(r)} \left( f(r) \left( \frac{1}{r \sqrt{T'(r)}} \right)' \right)' + \left( \frac{L(L+n-1)-1}{n-1} - 1 \right) \frac{T'(r)}{r T(r)} \right] \ .
\ede
 There exists no instability in this sector.

\subsubsection{Scalar-type perturbations}

Finally, the master equation for scalar perturbations takes the following form
\bna
V_{s}(r) &=& f(r) \left[ 2 \frac{L(L+n-1)}{n} \frac{(X(r) T(r) r)'}{X(r) T(r) r^{2}} - \frac{1}{X(r)} \left( f(r) X'(r) \right)' + 2 f(r) \left( \frac{X'(r)}{X(r)} \right)^{2} \right. \no \\
         && \hspace{2cm} \left. - \frac{1}{T(r)} \left( f(r) T'(r) \right)' + 2 f(r) \left( \frac{T'(r)}{T(r)} \right)^{2} + 2 f(r) \frac{X'(r) T'(r)}{X(r) T(r)} \right]
\ .
\ena
 In this case, it is also known that there exists the instability of small black holes under scalar perturbations in  odd dimensions.

The Lovelock black holes are unstable in various ways.
 The instability of black holes caused by the ghost is characterized by the violation of the $T'$-condition.
 As we mentioned, there are other types of instability under tensor and scalar type perturbations.
 The instability comes from the negative region of the effective potential. 
 For this class of black holes, we cannot use the WKB method because this method can be applicable to the positive definite effective potential with single peak.
 In this paper, we calculate the QNFs of stable black holes by excluding the cases with the negative regions or with the multipeaks in the potential.

\section{ Quasinormal Modes in Lovelock Gravity}

In this section, we present a novel method for the calculation of quasinormal modes in Lovelock gravity. 
 Quasinormal modes are  fundamental vibration modes around a black hole. 
 These modes are obtained by solving the master equation under the appropriate boundary condition.
 The general formalism of calculating QNFs by using WKB-approximation has been proposed by Schutz and Will~\cite{Schutz:1985zz} and subsequently developed by many people \cite{Iyer:1986np,Iyer:1986nq,Konoplya:2011qq}. 
 Here, we summarize the main points of the WKB-method for QNFs.

In general, we can divide the region into two regions. The region $I$ ($-\infty < x < x_0$) is the one ranging from top of the effective potential 
$x_0$ to the horizon of the black hole. 
 The region $II$ ($x_0 < x < \infty$) is the one ranging from top of the potential $x_0$ to the far outside of the black hole, i.e., infinity. 
 The wave traveling to the potential is called an ingoing wave and the wave traveling from the potential is called outgoing wave. In each region, the solutions of the master equation can be expressed as
\bee
\left\{ \begin{array}{c} \Phi_{I} = Z^{in}_{I}\Phi^{in}_{I} + Z^{out}_{I} \Phi^{out}_{I} \ , \\
                         \hspace{2mm} \\
                         \Phi_{II} = Z^{in}_{II}\Phi^{in}_{II} + Z^{out}_{II} \Phi^{out}_{II} \ , \end{array} \right. 
\ede
where $\Phi^{in}$ and  $\Phi^{out}$ represent ingoing and outgoing wave, respectively. 
The boundary condition for obtaining quasinormal modes is that there are no ingoing waves: 
\bee
Z^{in}_{I} = Z^{in}_{II}= 0 \ .
\ede
 Since there are two conditions, only discrete complex eigenvalues are allowed. 

In the $N$th-order WKB-method, we approximate the function $Q(x)$ defined by 
\bee
Q(x) \equiv \omega^{2} - V(x) .
\ede
in terms of $2N$th-order Taylor expansion around the maximum of the potential as follows;  
\bee
\left.\frac{\mathrm{d}Q(x)}{\mathrm{d}x} \right|_{x=x_{0}}=Q^{(1)}_{0} = 0 \hspace{5mm}{\rm and}\hspace{5mm} Q(x) \simeq \sum^{2N}_{p=0} \frac{1}{p!} Q^{(p)}_{0}( x - x_{0} )^{p} \ .
\ede
 Expressing the wave function using WKB-approximation, we can calculate the scattering matrix.
 Thus, we can get the formula for QNFs as
\bee
\omega \simeq \sqrt{ V_{0} + \sqrt{\frac{V^{(2)}_{0}}{2}} \left( n_{{\rm tone}} + \frac{1}{2} + \sum^{N-1}_{p=0} \Omega_{p} \right) } \ ,
\ede
where $V^{(p)}_0$ is the $p$th order derivative of the potential
\bee
V^{(p)}_0 \equiv \left.\frac{\mathrm{d}^{p}V(x)}{\mathrm{d}x^{p}} \right|_{x=x_{0}} \ ,
\ede
and the first and the second of $\Omega_{p}$ are given by
\bna
\Omega_1 &=& -30\,\left( n_{{\rm tone}} + \frac{1}{2} \right)^{2}{\beta_1}^{2} + 6\,\beta_2\,\left( n_{{\rm tone}} + \frac{1}{2} \right)^{2}-7/2\,{\beta_1}^{2}+3/2\,\beta_2, \\
       && \no \\
\Omega_2 &=& \left( n_{{\rm tone}} + \frac{1}{2} \right)^{3} \left( -2820\,{\beta_1}^{4} + 1800 {\beta_1}^{2}\beta_2 - 280\,\beta_1\,\beta_3 - 68\,{\beta_2}^{2} + 20\,\beta_4 \right) \no \\
&& \,\,\,\,\,\,\,\, + \left( n_{{\rm tone}} + \frac{1}{2} \right) \left( -1155\,{\beta_1}^{4}  + 918\,{\beta_1}^{2}\beta_2 - 190\,\beta_1\,\beta_3 - 67\,{\beta_2}^{2} + 25\,\beta_4 \right) \ .
\ena
 Here, $\beta_{p}\,\,(p\geq 1)$ is defined as
\bee
\beta_{p}\equiv \frac{V^{(p+2)}_{0}}{(p+2)!} \left( \frac{1}{2 V^{(2)}_{0}} \right)^{\dis \frac{p}{4}+1}
\ede
 The parameter, $n_{{\rm tone}}$, is called tone number of quasinormal modes.
 This method is often called the $N$th-order-WKB-approximation.
 This time, we used the 3rd-order WKB method. So we need $\Omega_{1}$ and $\Omega_{2}$.
 It is known that in the case of $n_{{\rm tone}} < L$ this approximation is good.
 So we focused on the case $n_{{\rm tone}}=0$ in this paper.

\subsection{WKB method adapted to Lovelock gravity}

Let us calculate QNFs of the fields in Lovelock gravity semianalytically by using the WKB method. 
 To use the WKB method, we need to know the potential function and hence the metric function $f(r)$.
 That is, we need to solve the $k$th order algebraic equation  (\ref{algebraic_eq}) for $\psi(r)$ in terms of $r$. 
 Unfortunately, it is impossible to solve the algebraic equation analytically for arbitrary dimensions. Thus, we need to invent a novel method adapted to Lovelock gravity.

To circumvent this difficulty, we take the following strategy. 
 Instead of solving Eq. $(\ref{algebraic_eq})$, we change the variable from $r$ to $\psi$ using the relation
\bee
r = \sqrt[n+1]{\frac{\mathcal{M}}{\mathcal{P}(\psi)}} \ .
\ede
 Thus, we are able to get the analytic representation of 
$f(\psi)$ as 
\bee
f = 1 - r^{2} \psi = 1 - \sqrt[\frac{n+1}{2}]{\frac{\mathcal{M}}{P(\psi)}} \psi \ . \label{f} 
\ede
 So, if we use this formulation, we will be able to calculate QNFs of the static and spherically symmetric black holes in Lovelock gravity.

Now, we check the range of $\psi$. 
 We are considering the static spherically symmetric  black hole solutions with the asymptotically flat region.
 When $r$ goes to $\infty$, $f(r)$ should be 1 because of asymptotic flatness. 
 Therefore, in the limit  $r \to \infty$, we have
\bee
\psi \to 0 \ .
\ede
 When $r$ approaches the horizon, $f(r)$ must vanish. 
 Hence, in this limit, we obtain
\bee
\left( \frac{P(\psi)}{\mathcal{M}} \right)^{2} = \psi^{n+1} \label{cond} \ .
\ede
 There will be the unique real solution $\psi_{\mathrm{h}}$ of this equation because we know $\mathcal{P}(\psi)$ is the monotonic increasing function of $\psi$. 
 To summarize, we obtained the relation between $r$ and $\psi$ as follow:
\bee
\left\{
\begin{array}{cccc}
 r: & horizon & \lr & \infty \\ 
 \psi: & \psi_{{\rm h}} & \lr & 0 
\end{array} .
\right.
\ede
 In this region of $\psi$, we must find $\psi_0$ which satisfy the equation,
\bee
V^{(1)}_{0} = 0 \ .
\ede
 Once we find it, we can use the WKB method and calculate QNFs of Lovelock black holes.

\begin{figure}[htbp]
  \begin{center}
   \caption{An example of good potentials.}
     \includegraphics[clip, height=5cm,bb=0 0 370 230]{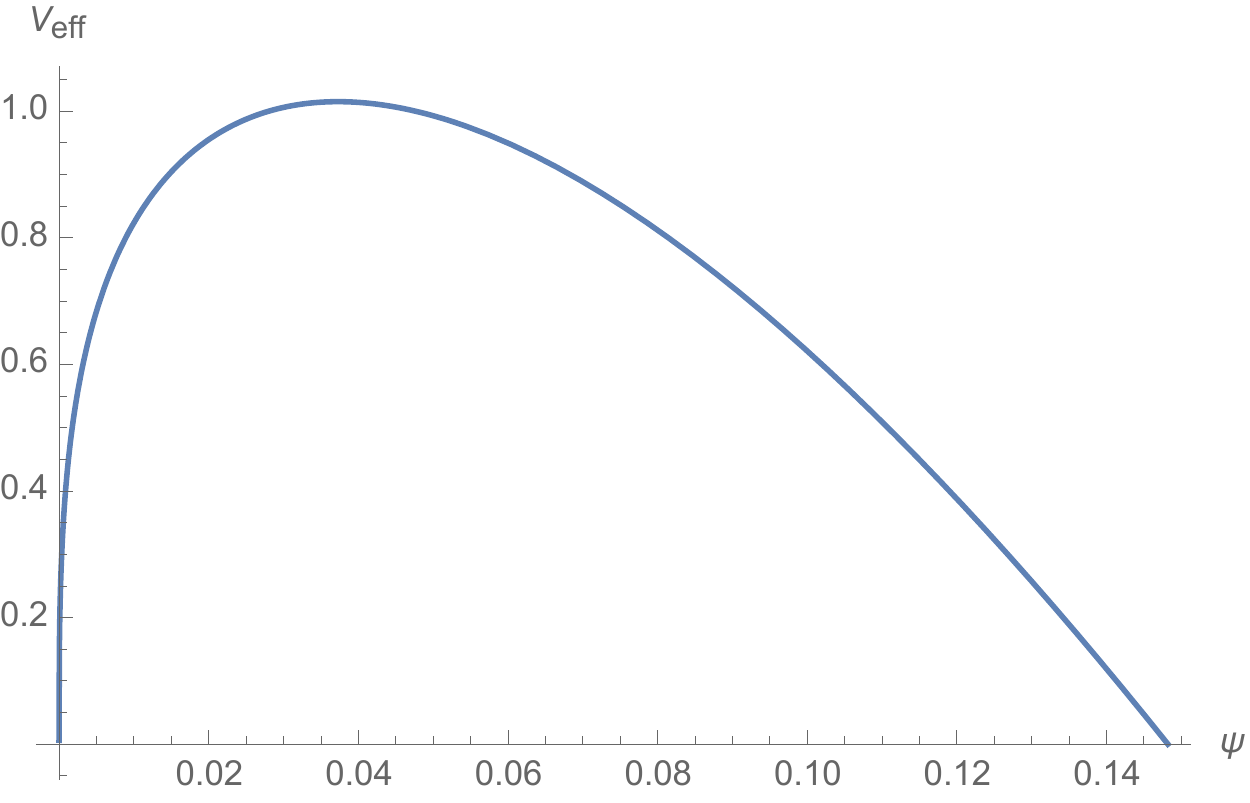} \\
     {\scriptsize $D=7$, $\mathcal{M}=50$, $L=2$, $A_{2}=0.65$, $A_{3}=0.1$ \\ the effective potential of tensor perturbation in tensor field}
     \label{GP}
  \end{center}
\end{figure}

\begin{figure}[htbp]
  \begin{center}
  \caption{Examples of bad potentials.}
    \begin{tabular}{c}

      % 1
      \begin{minipage}{0.5\hsize}
        \begin{center}
           {\scriptsize ex.1) double peaks}
          \includegraphics[clip, height=5cm,bb=0 0 360 230]{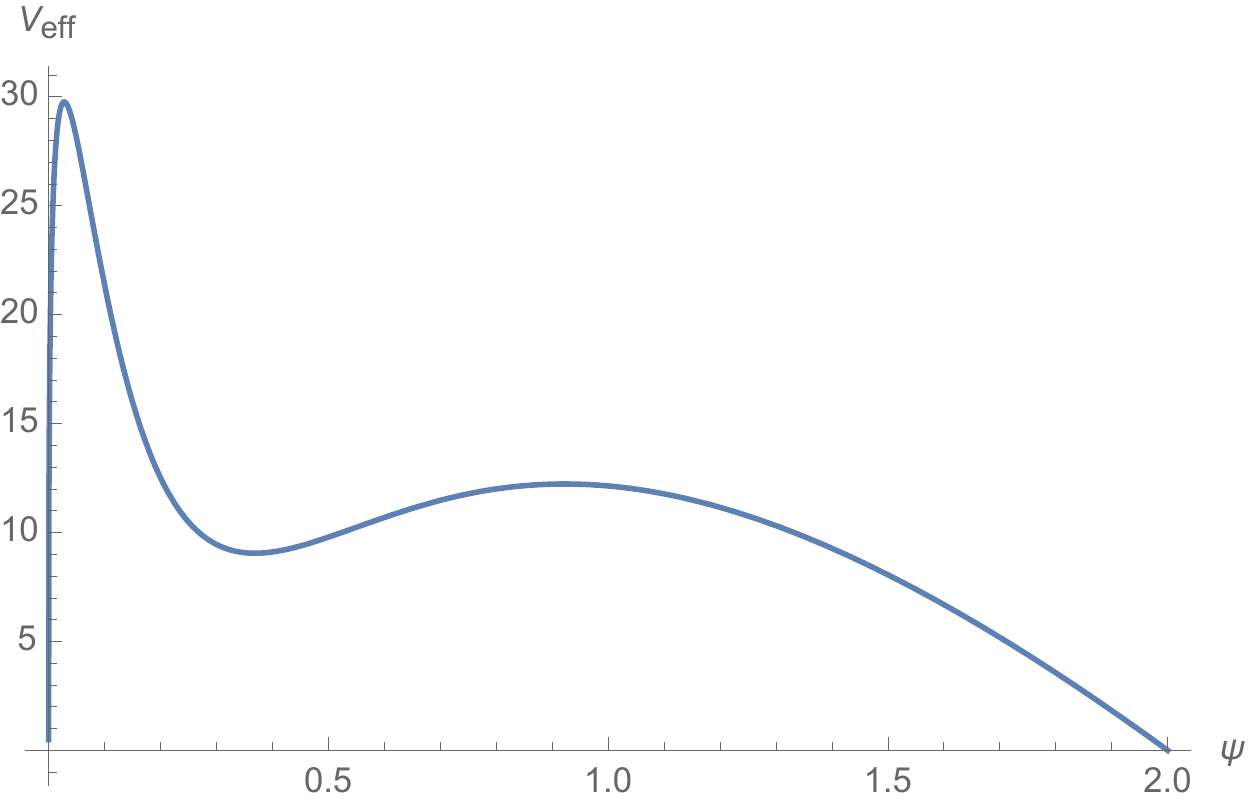}
          \hspace{1.6cm} {\scriptsize $D=7$, $\mathcal{M}=1$, $L=10$, $A_{2}=1.50$, $A_{3}=0.0$}
        \end{center}
      \end{minipage}

      % 2
      \begin{minipage}{0.5\hsize}
        \begin{center}
           {\scriptsize ex.2) negative region}
          \includegraphics[clip, height=5cm,bb=0 0 360 230]{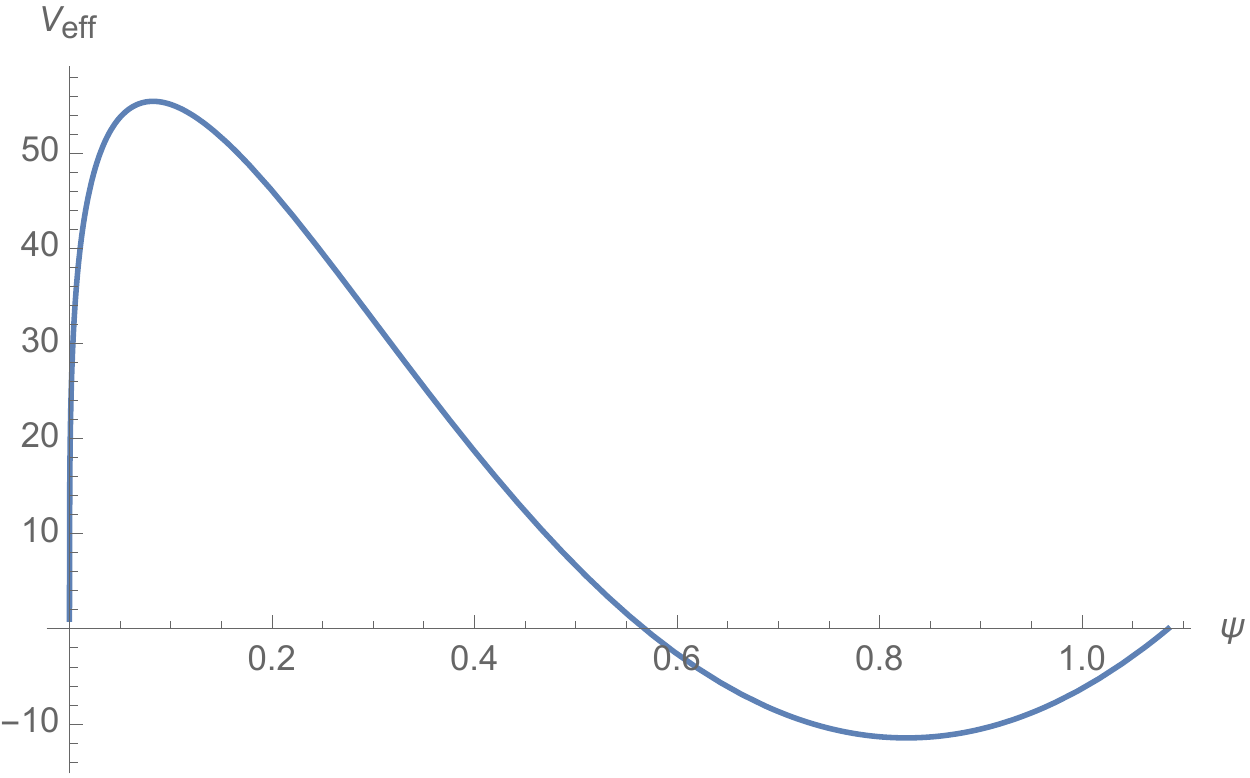}
          {\scriptsize $D=8$, $\mathcal{M}=1$, $L=10$, $A_{2}=0.05$, $A_{3}=0.1$}
        \end{center}
      \end{minipage}

    \end{tabular}
    \label{NRDP}\\
    {\scriptsize the effective potential of scalar perturbation in tensor field}
  \end{center}
\end{figure}

\subsection{Numerical results}
In this subsection, we calculate the QNFs of Lovelock black holes in seven  and eight dimensions using the WKB method adapted to Lovelock gravity.

 From now on, we rewrite the parameters as
\bee
A_{q}\equiv\left\{ \begin{array}{l} 1 \hspace{2.85cm}(q=1) \ , \\
                                    \\
                                    \dis \frac{a_q \prod^{2q-2}_{p=1}(n-p)}{q}\hspace{5mm}(q\geq2) \ .
                                    \end{array} \right.
\ede
We calculated the QNFs for various values of $A_{q}(q\geq 2)$ using the third order WKB method.
 Note that we fixed the $\Lambda$ parameter to zero and considered a  massless scalar field ($\mu=0$). 
 We calculated QNFs only for the cases like in Fig.\ref{GP}.
 Namely, the potential is positive definite in the range $0<\psi<\psi_{\rm h}$ and has a single peak.
 For some parameters, the effective potential has  the multipeaks or the negative region  (Fig.\ref{NRDP}).
 We excluded these and the cases with ghost instability from the calculation.

\begin{figure}[htbp]
  \begin{center}
  \caption{The quasinormal mode diagram of scalar field $L=0$}
    \begin{tabular}{c}
      % 3
      \begin{minipage}{0.5\hsize}
        \begin{center}
          \includegraphics[clip, width=7.5cm,bb=0 0 1024 768]{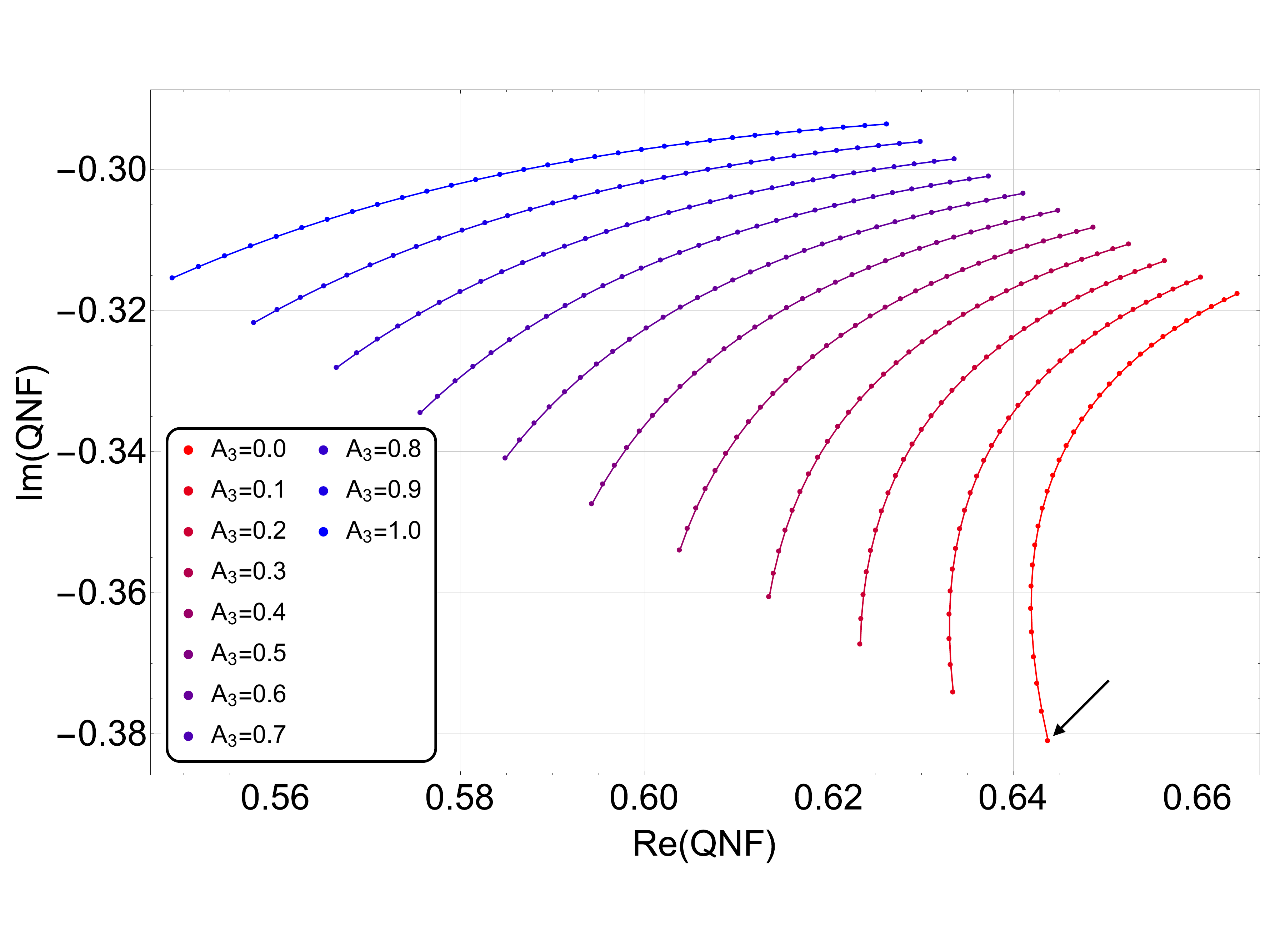}
          \hspace{1.6cm} {\scriptsize $D=7,\hspace{2mm}\mathcal{M}=10$}
        \end{center}
      \end{minipage}

      % 4
      \begin{minipage}{0.5\hsize}
        \begin{center}
          \includegraphics[clip, width=7.5cm,bb=0 0 1024 768]{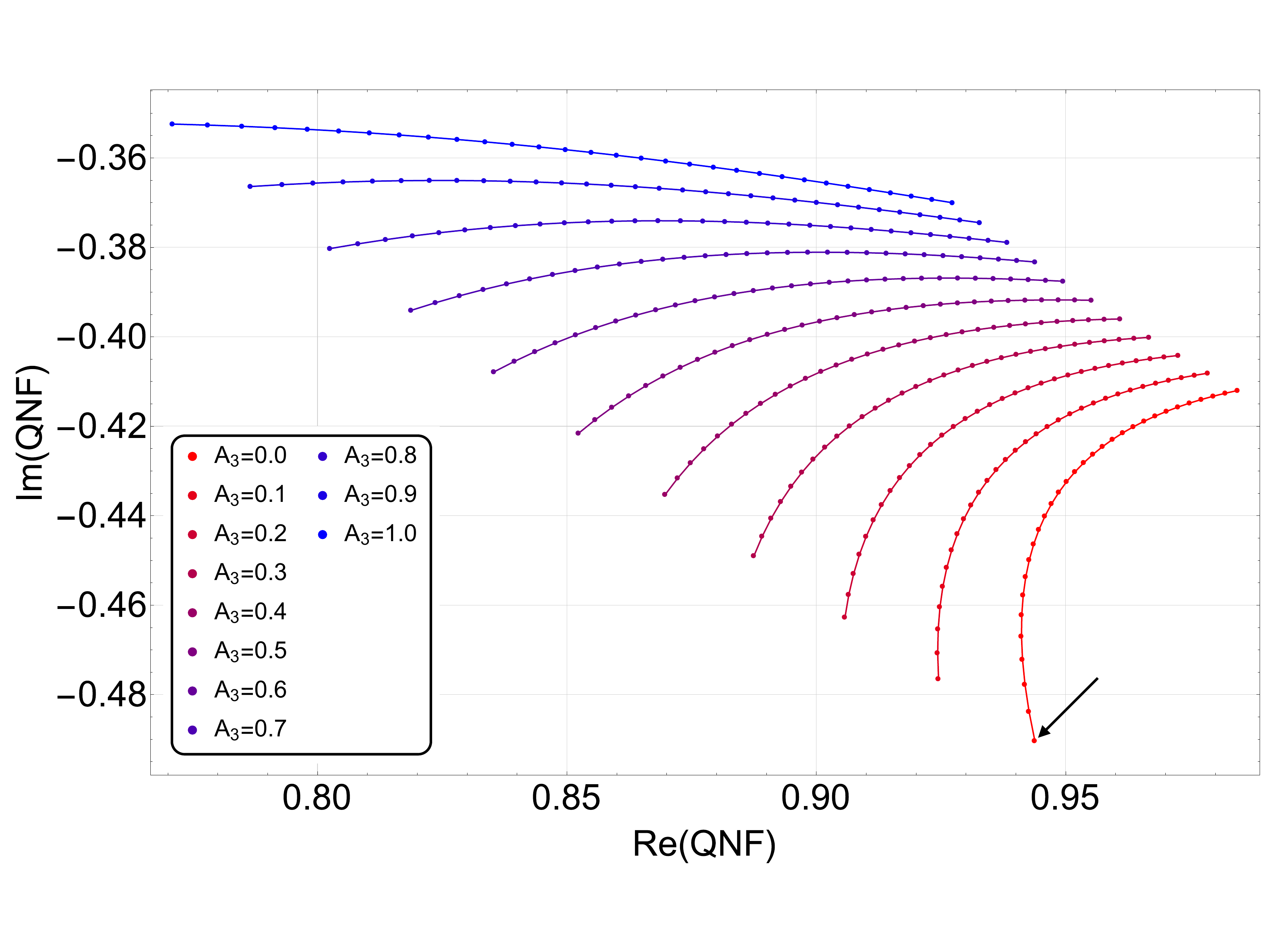}
          \hspace{1.6cm} {\scriptsize $D=8,\hspace{2mm}\mathcal{M}=10$}
        \end{center}
      \end{minipage}
      
      \\
      \\
      \\

      % 5
      \begin{minipage}{0.5\hsize}
        \begin{center}
          \includegraphics[clip, width=7.5cm,bb=0 0 1024 768]{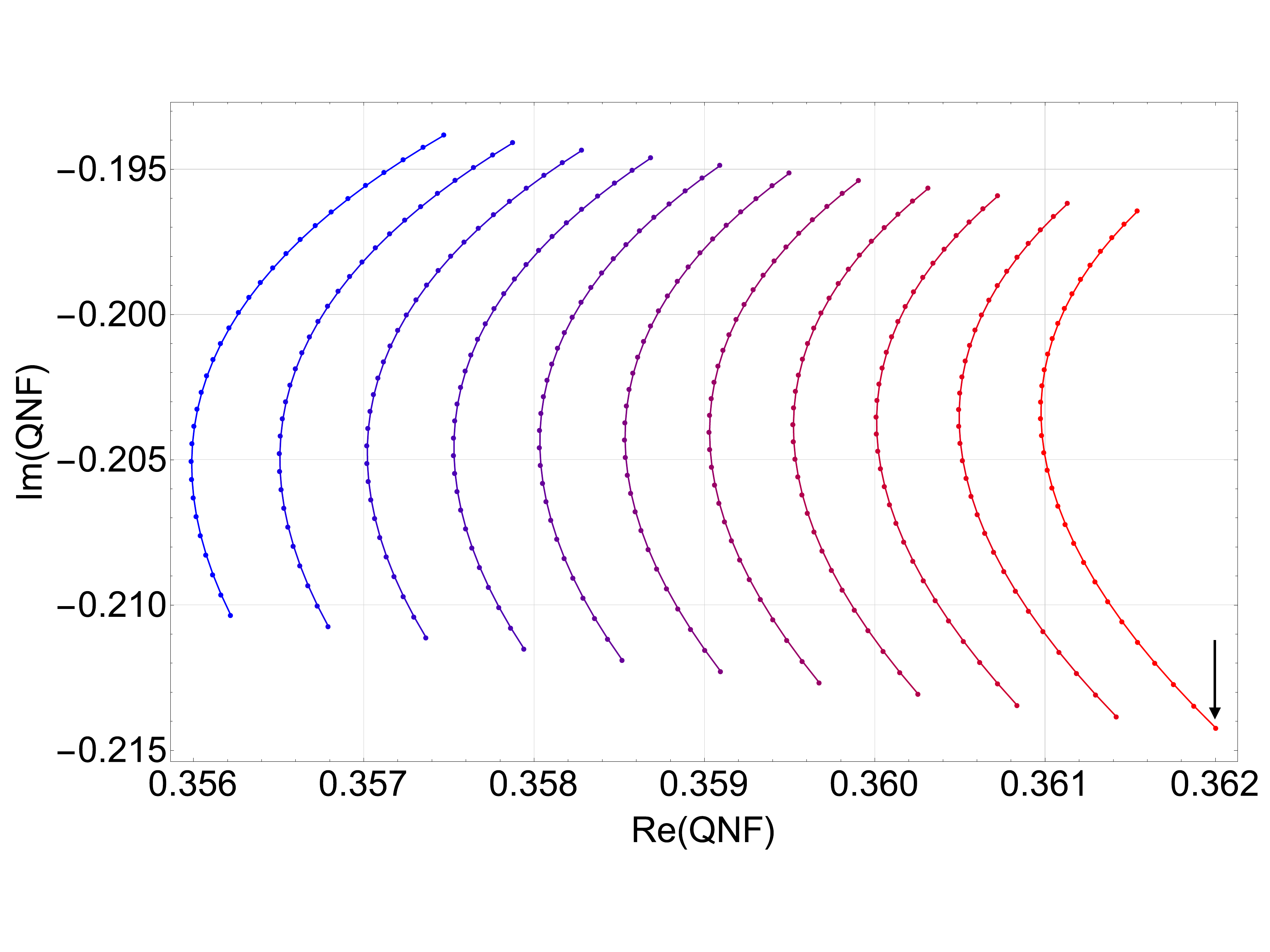}
          \hspace{1.6cm} {\scriptsize $D=7,\hspace{2mm}\mathcal{M}=100$}
        \end{center}
      \end{minipage}

      % 6
      \begin{minipage}{0.5\hsize}
        \begin{center}
          \includegraphics[clip, width=7.5cm,bb=0 0 1024 768]{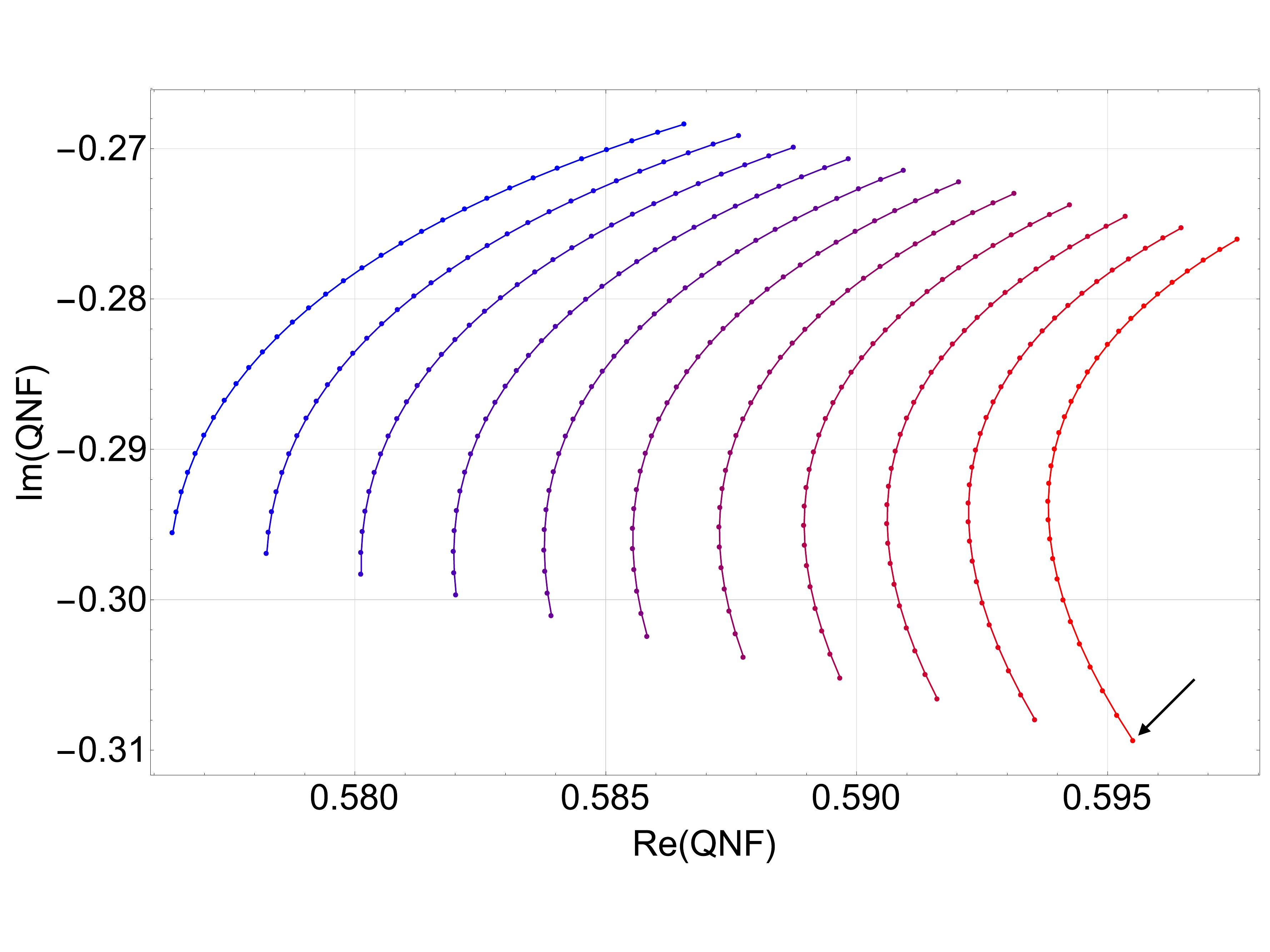}
          \hspace{1.6cm} {\scriptsize $D=8,\hspace{2mm}\mathcal{M}=100$}
        \end{center}
      \end{minipage}
      
      \\
      \\
      \\

      % 7
      \begin{minipage}{0.5\hsize}
        \begin{center}
          \includegraphics[clip, width=7.5cm,bb=0 0 1024 768]{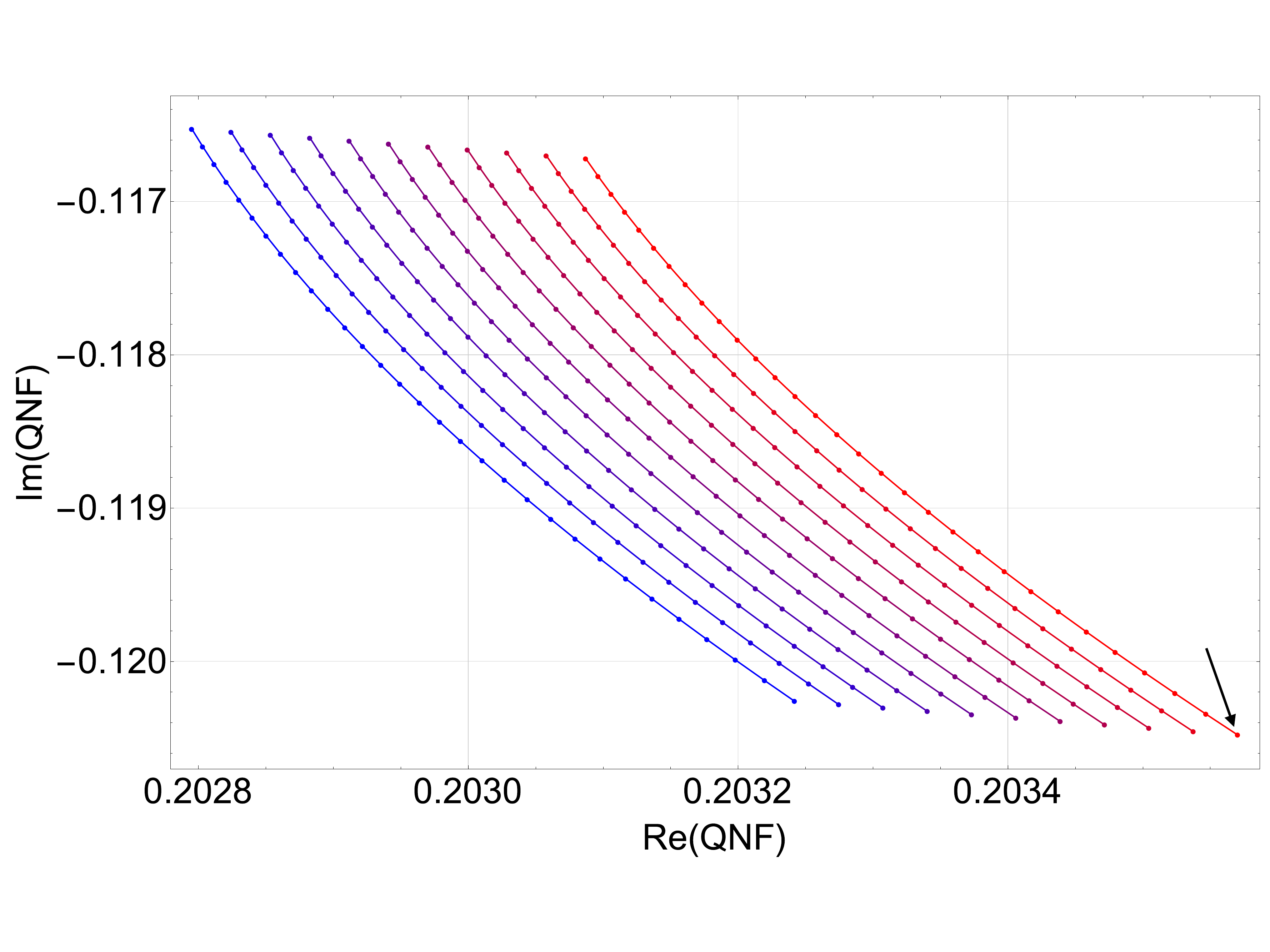}
          \hspace{1.6cm} {\scriptsize $D=7,\hspace{2mm}M=1000$}
        \end{center}
      \end{minipage}

      % 8
      \begin{minipage}{0.5\hsize}
        \begin{center}
          \includegraphics[clip, width=7.5cm,bb=0 0 1024 768]{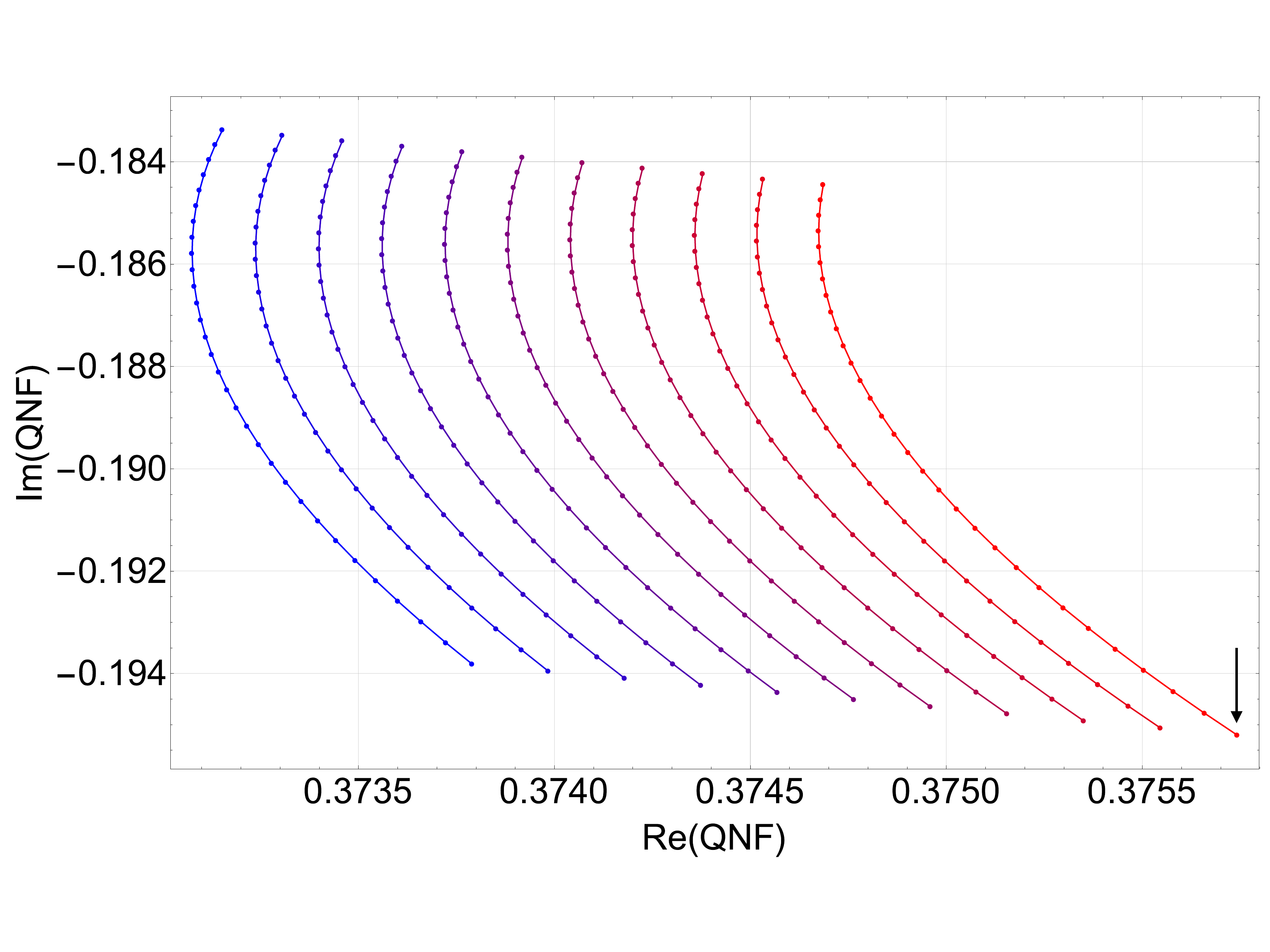}
          \hspace{1.6cm} {\scriptsize $D=8,\hspace{2mm}M=1000$}
        \end{center}
      \end{minipage}
    
    \end{tabular}
    \label{SF}
  \end{center}
\end{figure}

\begin{figure}[htbp]
  \begin{center}
  \caption{The quasinormal mode diagram of tensor perturbation in tensor field $L=2$}
    \begin{tabular}{c}

      % 1
      \begin{minipage}{0.5\hsize}
        \begin{center}
          \includegraphics[clip, width=7.5cm,bb=0 0 1024 768]{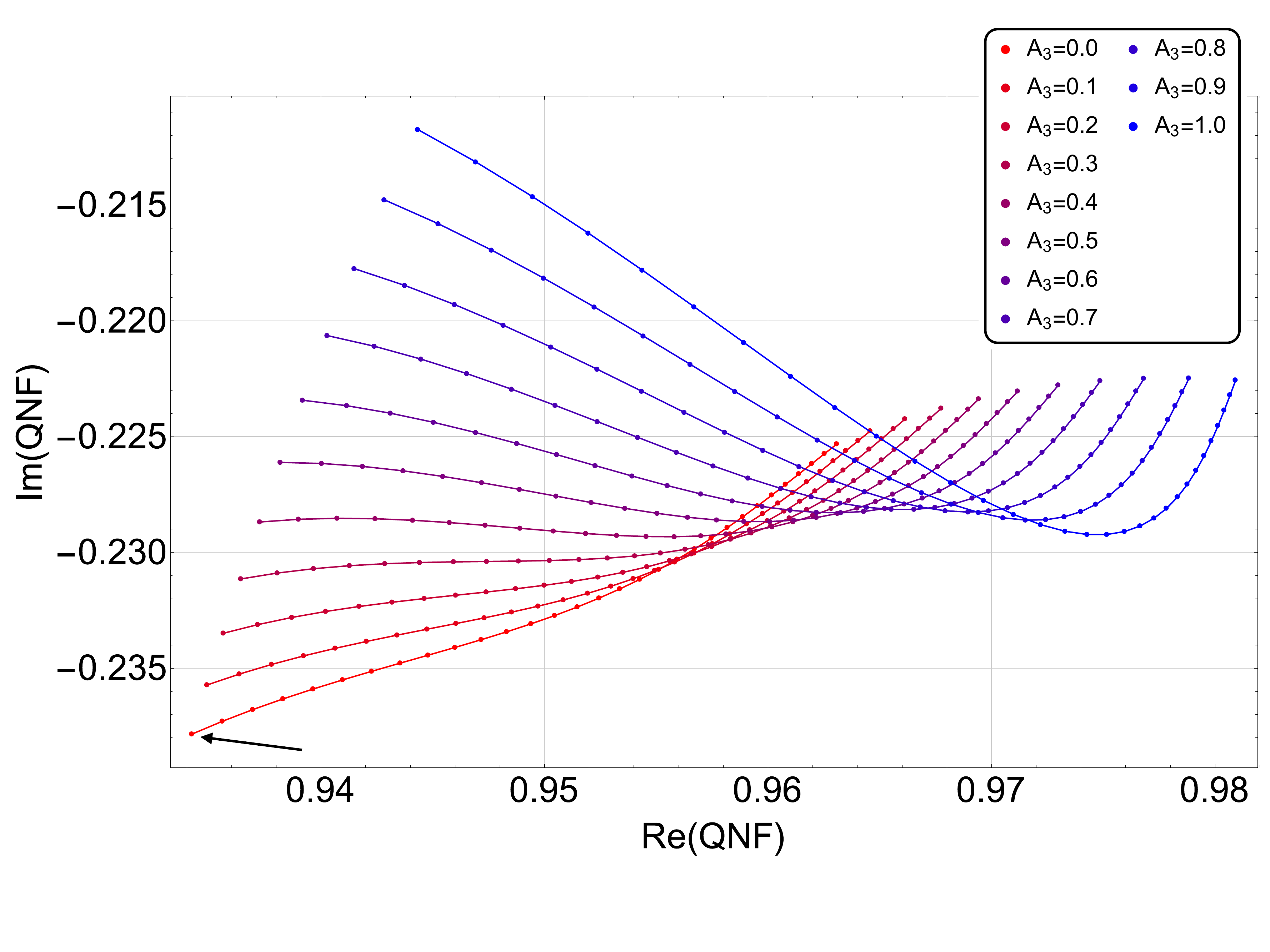}
          \hspace{1.6cm} {\scriptsize $D=7,\hspace{2mm}\mathcal{M}=50$}
        \end{center}
      \end{minipage}

      % 2
      \begin{minipage}{0.5\hsize}
        \begin{center}
          \includegraphics[clip, width=7.5cm,bb=0 0 1024 768]{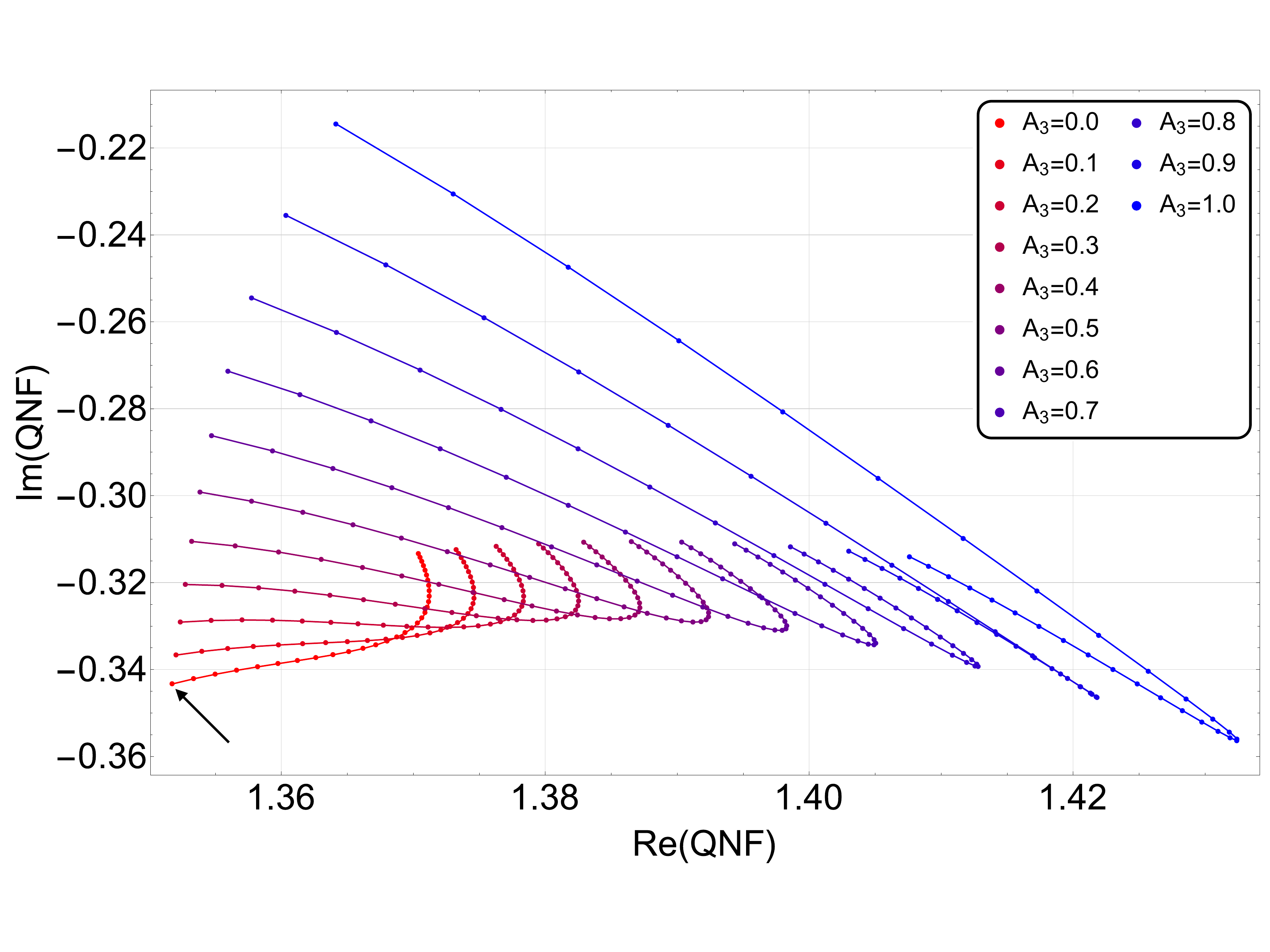}
          \hspace{1.6cm} {\scriptsize $D=8,\hspace{2mm}\mathcal{M}=50$}
        \end{center}
      \end{minipage}
      
      \\
      \\
      \\

      % 3
      \begin{minipage}{0.5\hsize}
        \begin{center}
          \includegraphics[clip, width=7.5cm,bb=0 0 1024 768]{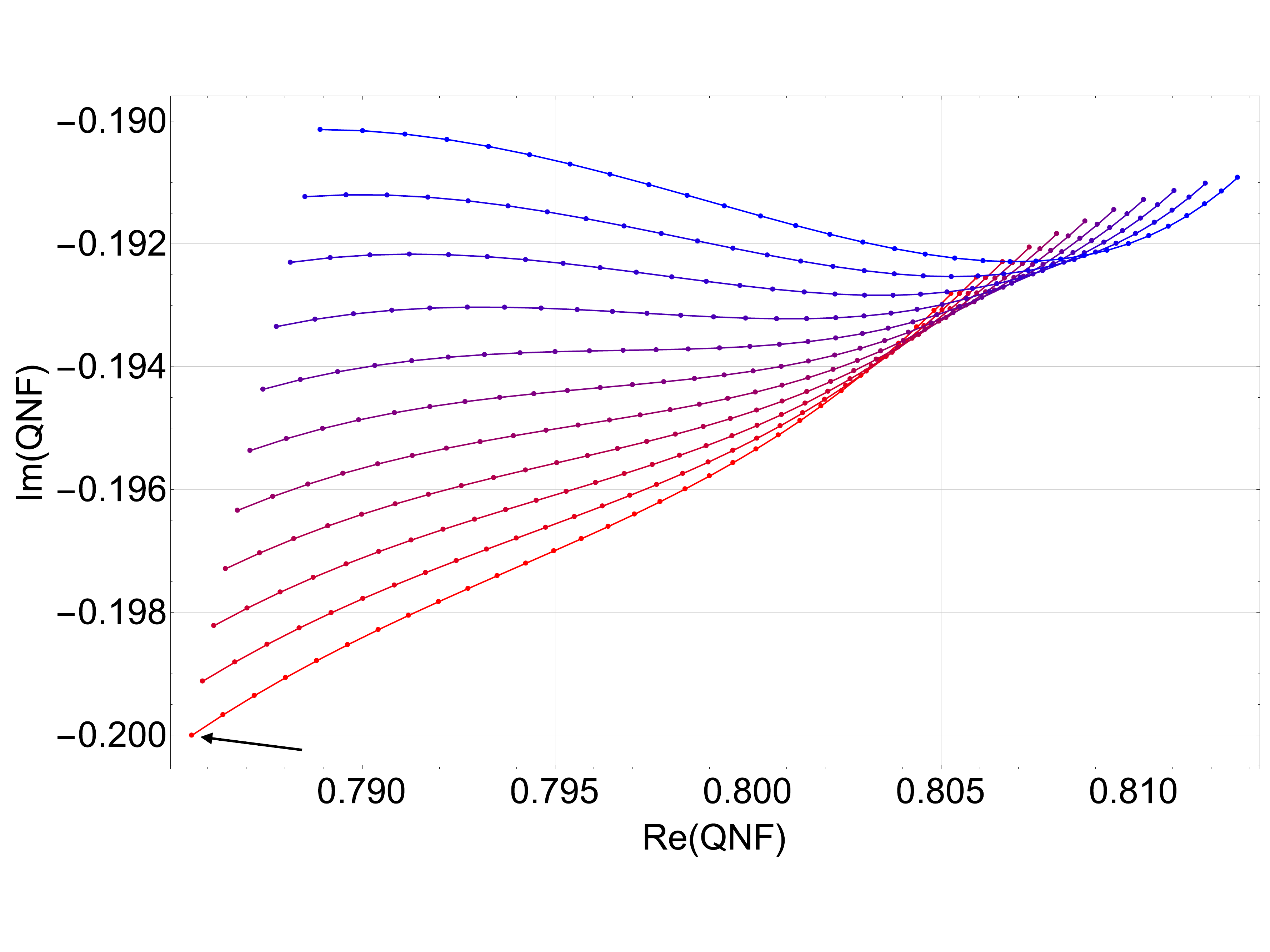}
          \hspace{1.6cm} {\scriptsize $D=7,\hspace{2mm}M=100$}
        \end{center}
      \end{minipage}

      % 4
      \begin{minipage}{0.5\hsize}
        \begin{center}
          \includegraphics[clip, width=7.5cm,bb=0 0 1024 768]{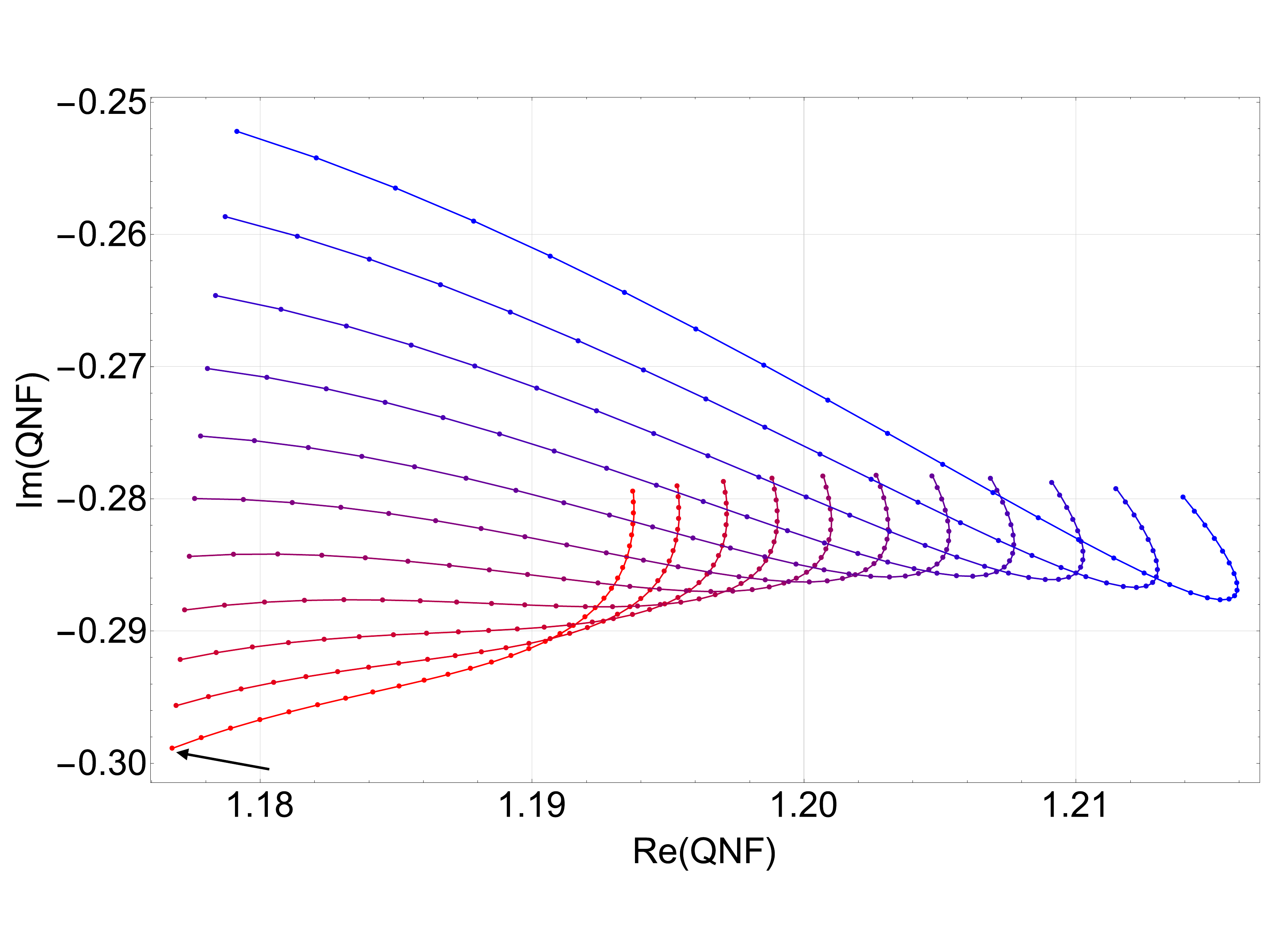}
          \hspace{1.6cm} {\scriptsize $D=8,\hspace{2mm}M=100$}
        \end{center}
      \end{minipage}

      \\
      \\
      \\
      
      % 5
      \begin{minipage}{0.5\hsize}
        \begin{center}
          \includegraphics[clip, width=7.5cm,bb=0 0 1024 768]{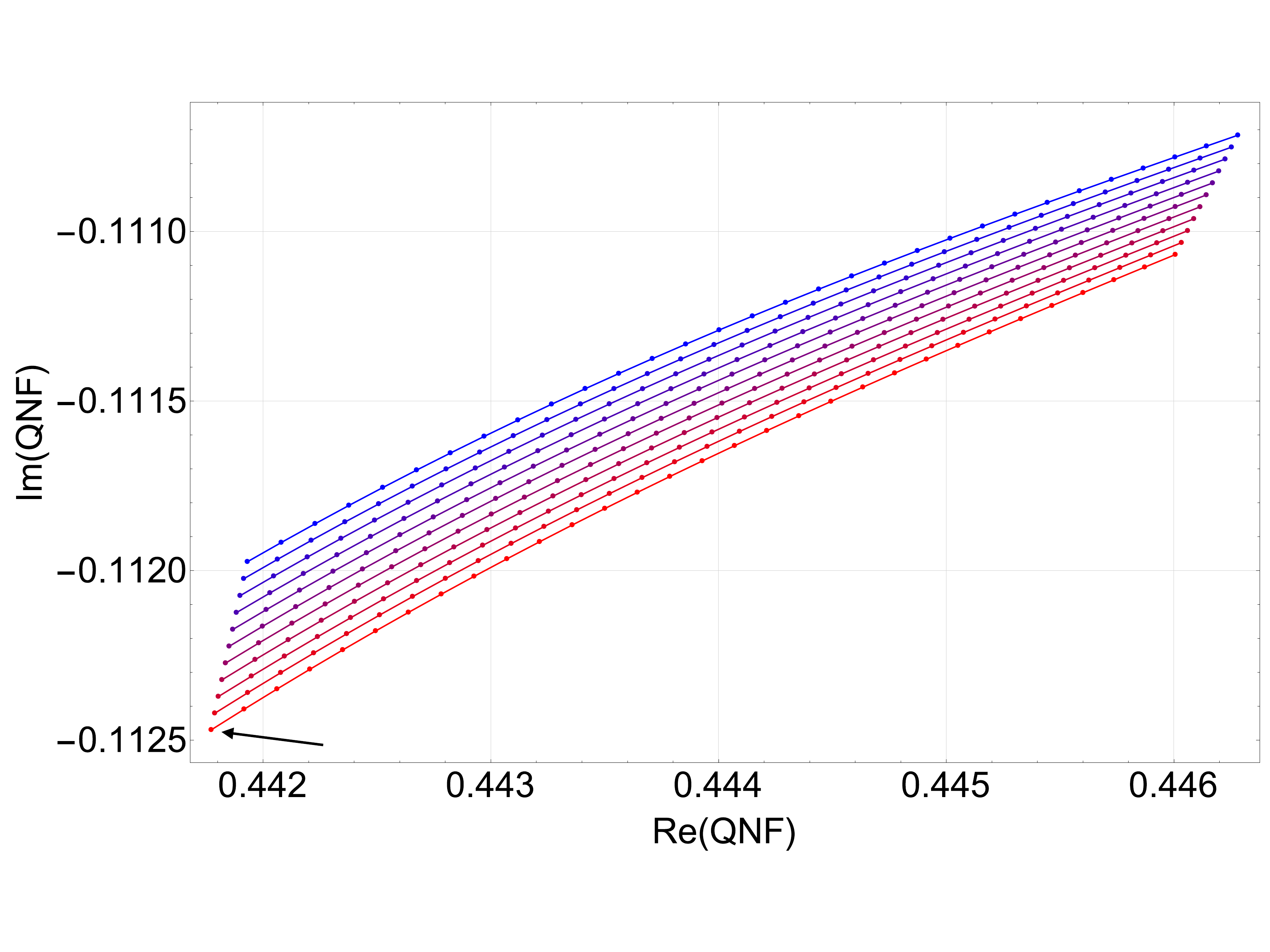}
          \hspace{1.6cm} {\scriptsize $D=7,\hspace{2mm}\mathcal{M}=1000$}
        \end{center}
      \end{minipage}

      % 6
      \begin{minipage}{0.5\hsize}
        \begin{center}
          \includegraphics[clip, width=7.5cm,bb=0 0 1024 768]{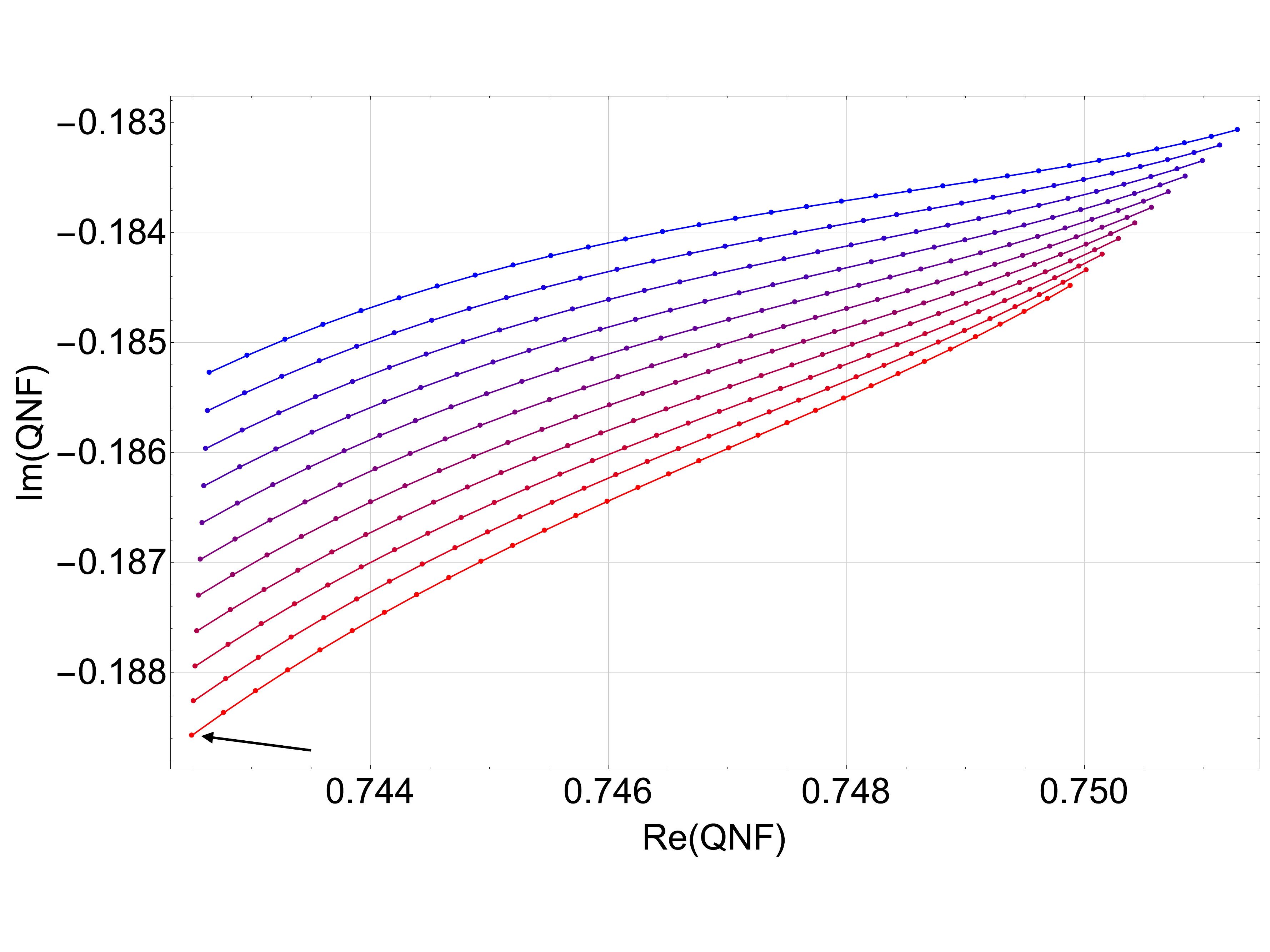}
          \hspace{1.6cm} {\scriptsize $D=8,\hspace{2mm}\mathcal{M}=1000$}
        \end{center}
      \end{minipage}

    \end{tabular}
    \label{TP}
  \end{center}
\end{figure}

\begin{figure}[htbp]
  \begin{center}
  \caption{The quasinormal mode diagram of vector perturbation in tensor field $L=10$}
    \begin{tabular}{c}

      % 1
      \begin{minipage}{0.5\hsize}
        \begin{center}
          \includegraphics[clip, width=7.5cm,bb=0 0 1024 768]{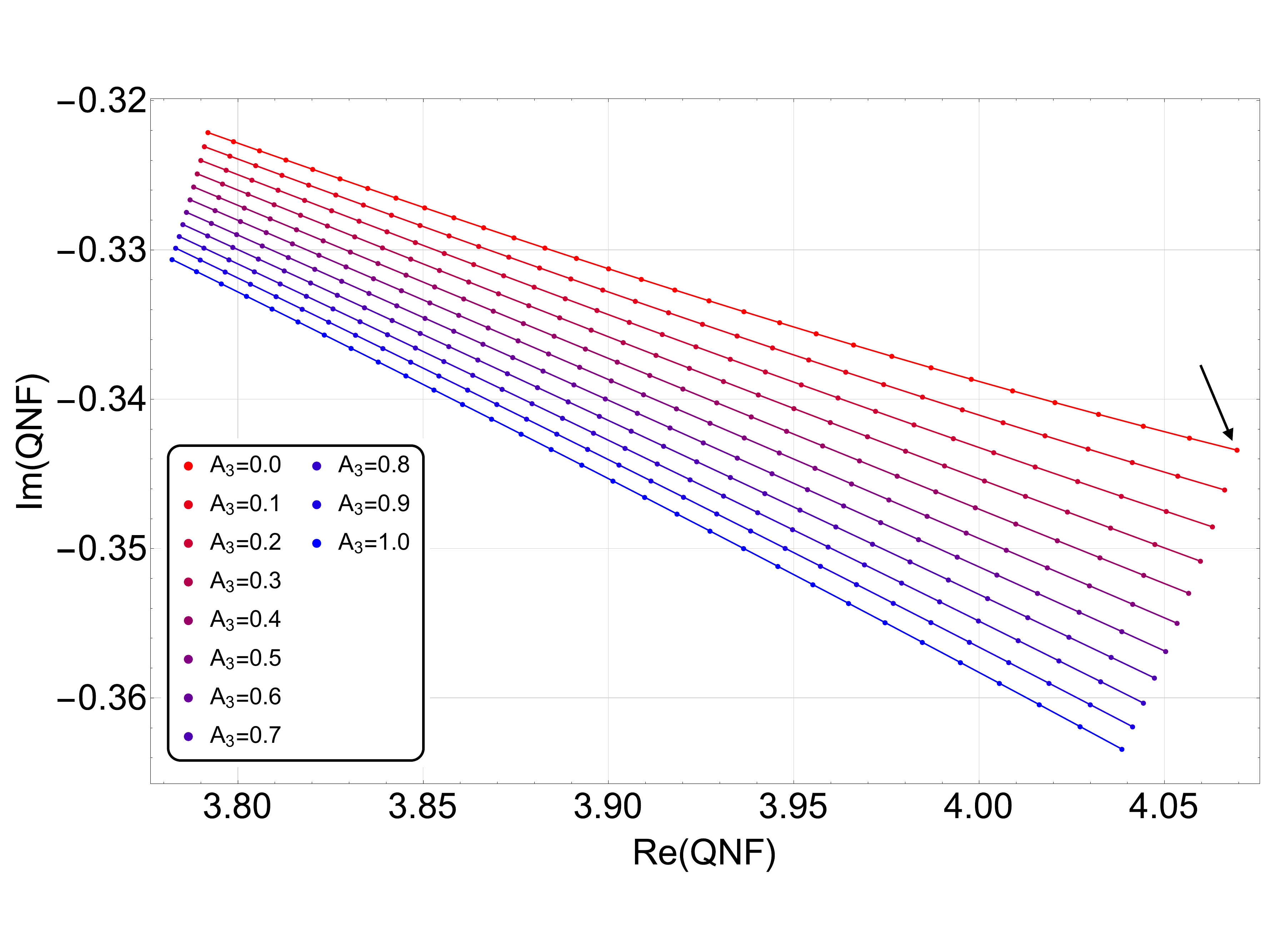}
          \hspace{1.6cm} {\scriptsize $D=7,\hspace{2mm}\mathcal{M}=10$}
        \end{center}
      \end{minipage}

      % 2
      \begin{minipage}{0.5\hsize}
        \begin{center}
          \includegraphics[clip, width=7.5cm,bb=0 0 1024 768]{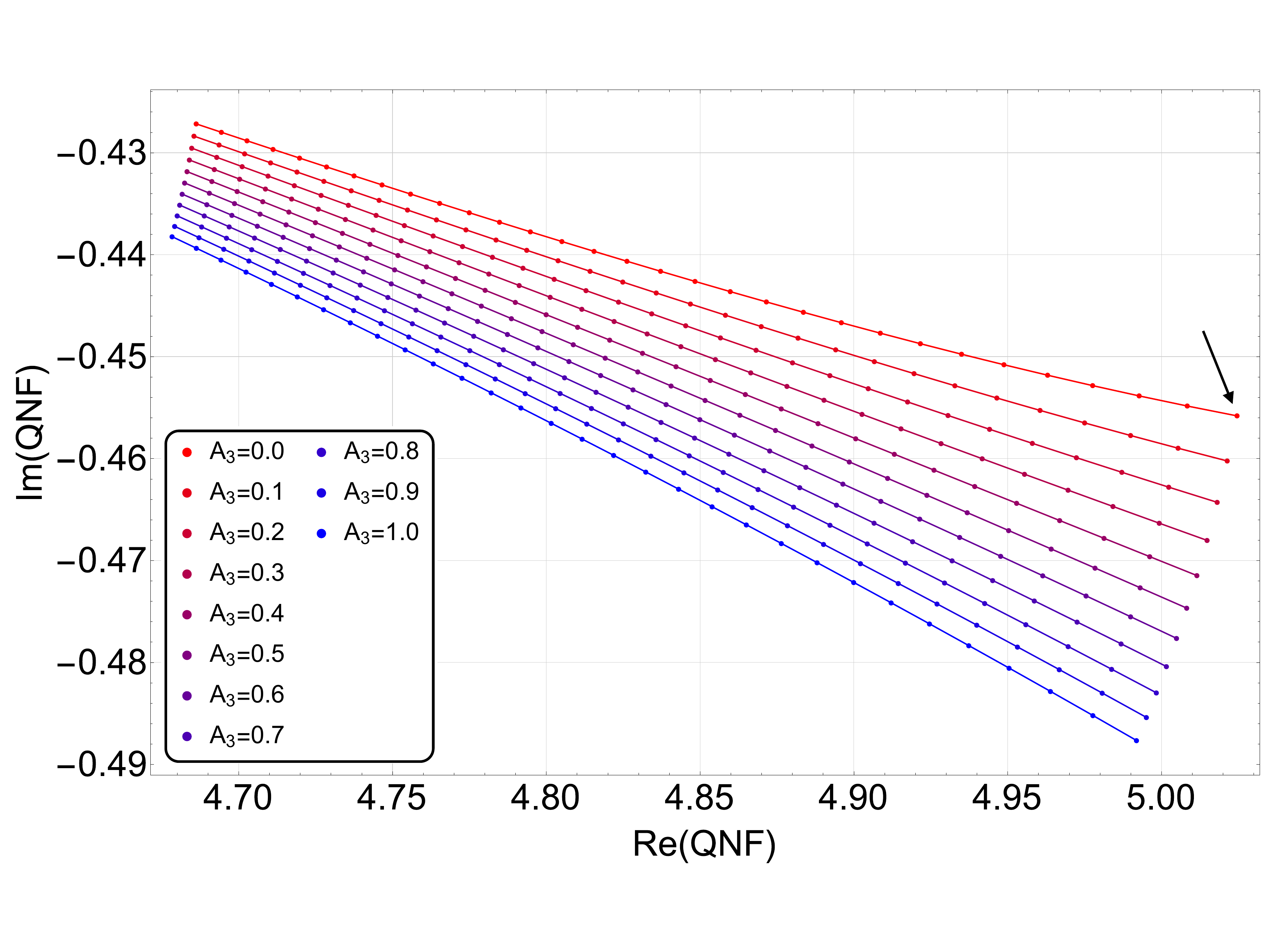}
          \hspace{1.6cm} {\scriptsize $D=8,\hspace{2mm}\mathcal{M}=10$}
        \end{center}
      \end{minipage}
      
      \\
      \\
      \\

      % 3
      \begin{minipage}{0.5\hsize}
        \begin{center}
          \includegraphics[clip, width=7.5cm,bb=0 0 1024 768]{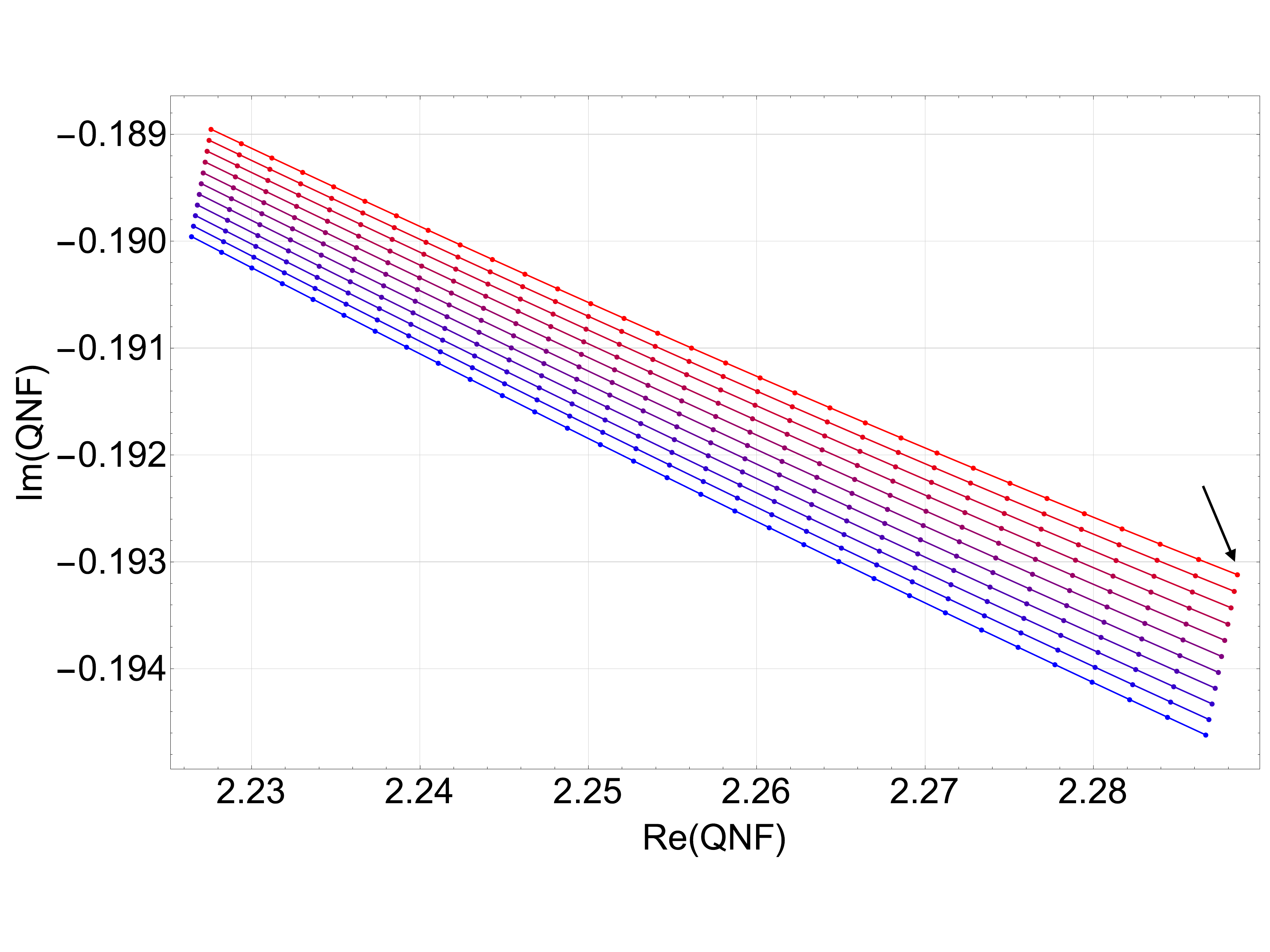}
          \hspace{1.6cm} {\scriptsize $D=7,\hspace{2mm}\mathcal{M}=100$}
        \end{center}
      \end{minipage}

      % 4
      \begin{minipage}{0.5\hsize}
        \begin{center}
          \includegraphics[clip, width=7.5cm,bb=0 0 1024 768]{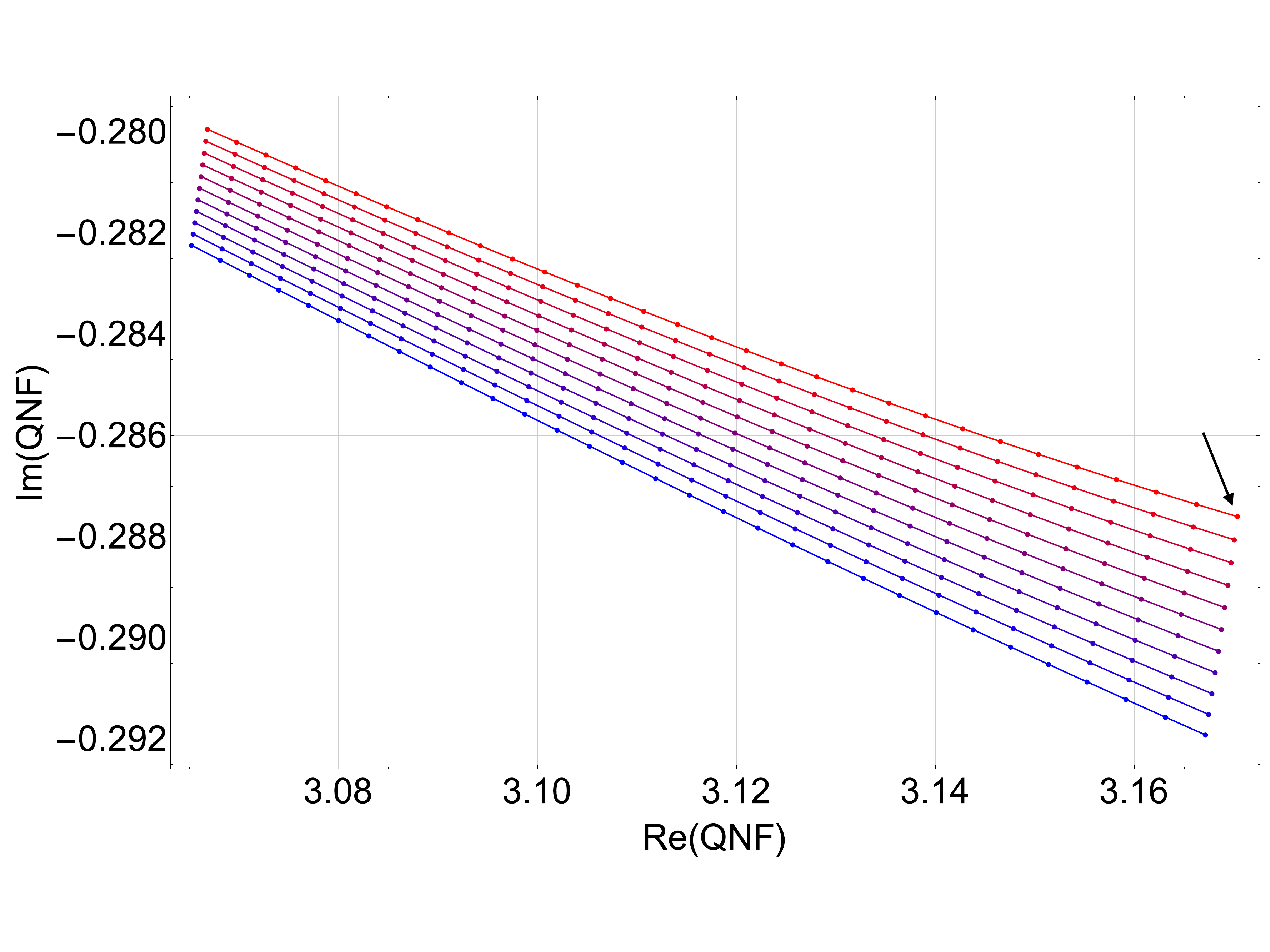}
          \hspace{1.6cm} {\scriptsize $D=8,\hspace{2mm}\mathcal{M}=100$}
        \end{center}
      \end{minipage}
      
      \\
      \\
      \\

      % 5
      \begin{minipage}{0.5\hsize}
        \begin{center}
          \includegraphics[clip, width=7.5cm,bb=0 0 1024 768]{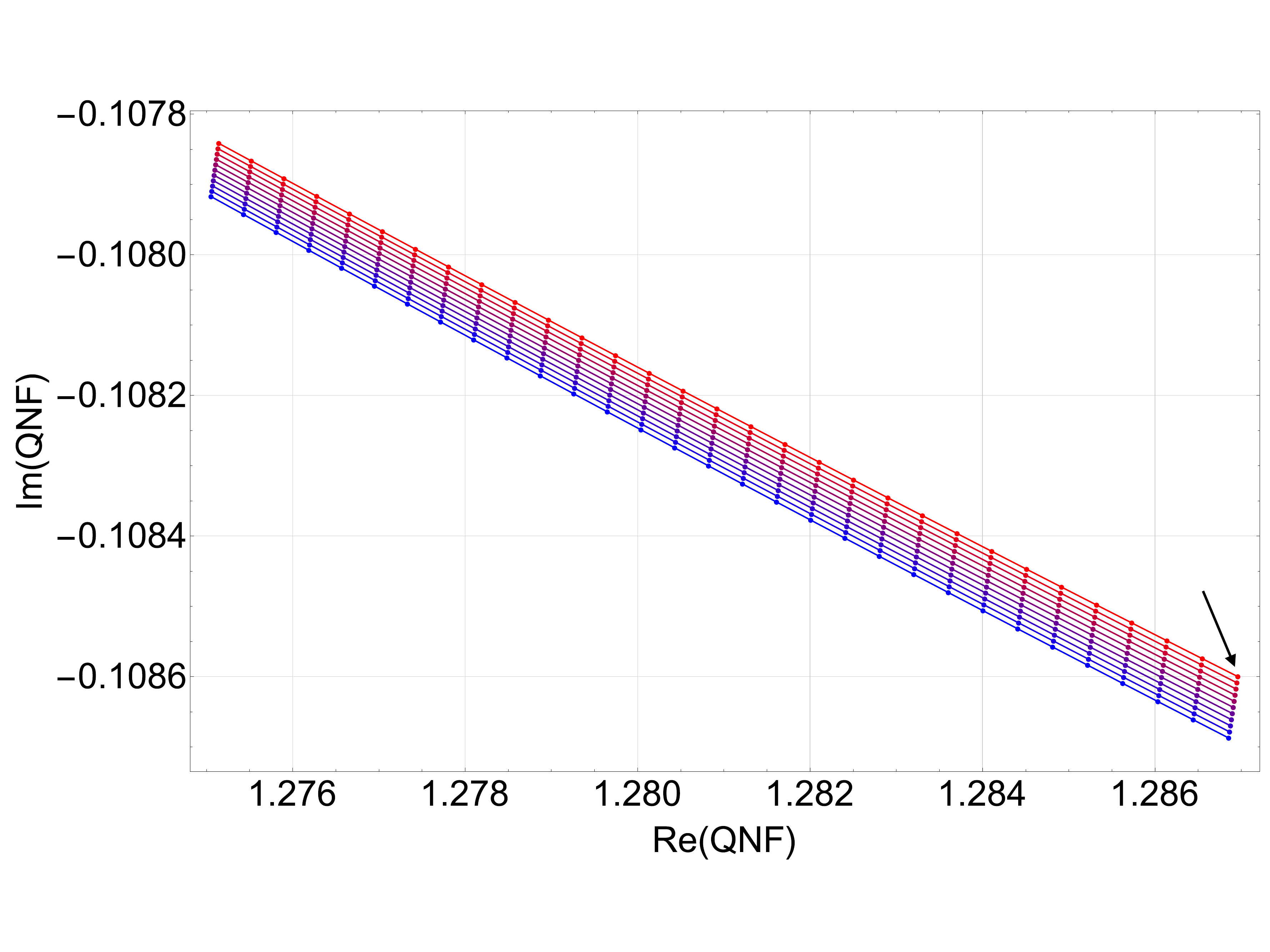}
          \hspace{1.6cm} {\scriptsize $D=7,\hspace{2mm}M=1000$}
        \end{center}
      \end{minipage}

      % 6
      \begin{minipage}{0.5\hsize}
        \begin{center}
          \includegraphics[clip, width=7.5cm,bb=0 0 1024 768]{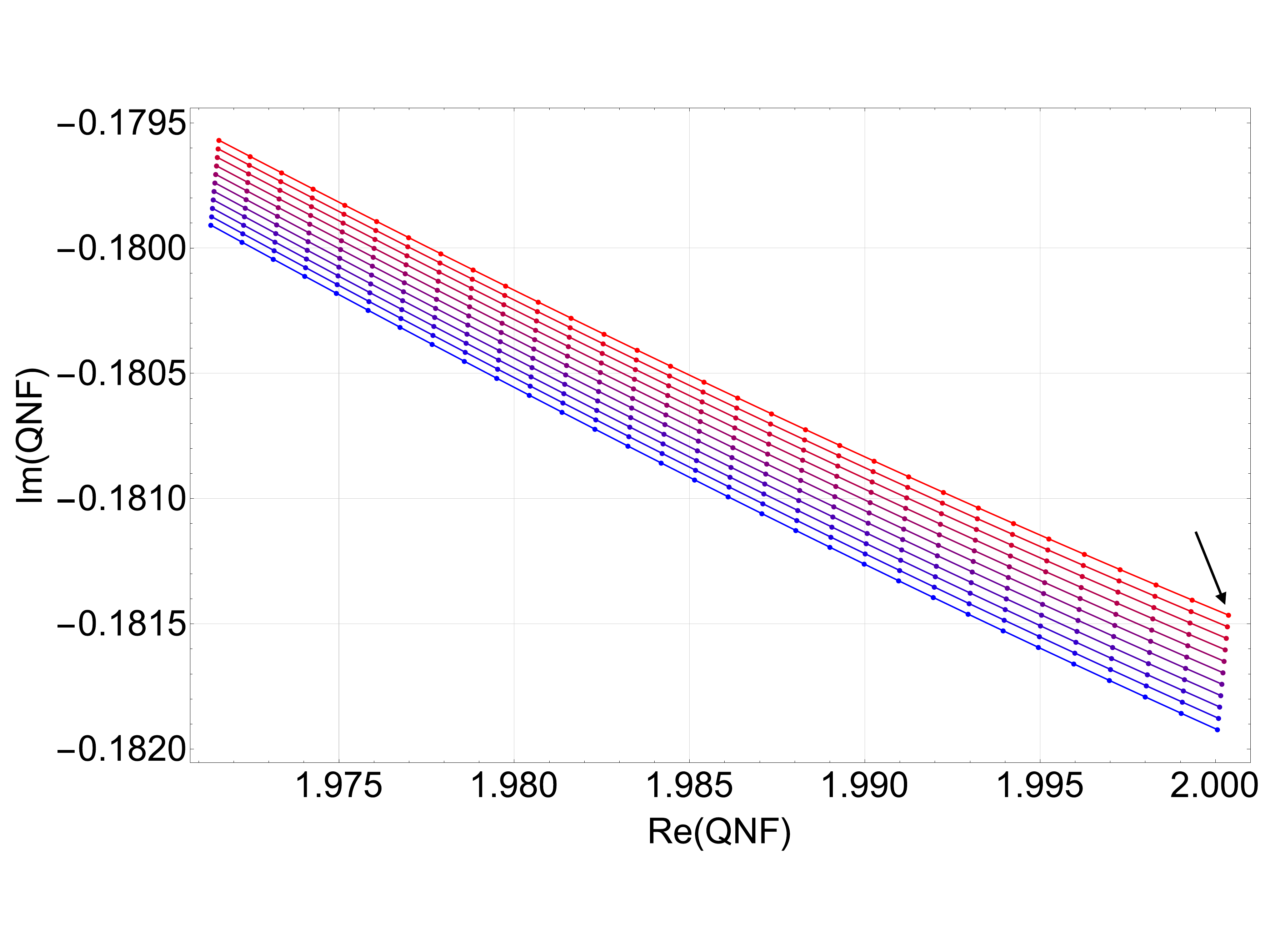}
          \hspace{1.6cm} {\scriptsize $D=8,\hspace{2mm}M=1000$}
        \end{center}
      \end{minipage}

    \end{tabular}
    \label{VP}
  \end{center}
\end{figure}

\begin{figure}[htbp]
  \begin{center}
      \caption{The quasinormal mode diagram of scalar perturbation in tensor field $L=10$}
    \begin{tabular}{c}

      % 1
      \begin{minipage}{0.5\hsize}
        \begin{center}
          \includegraphics[clip, width=7.5cm,bb=0 0 1024 768]{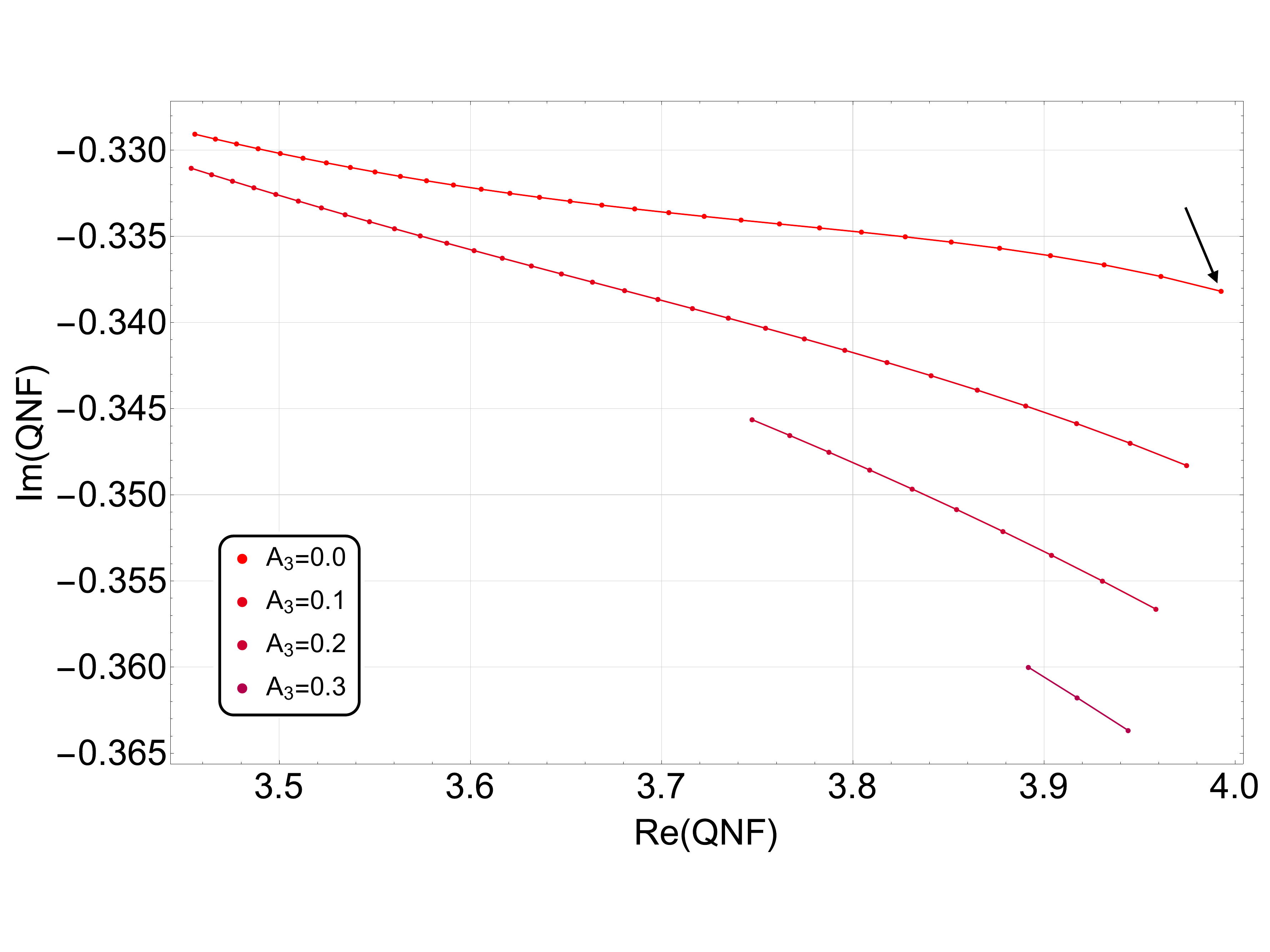}
          \hspace{1.6cm} {\scriptsize $D=7,\hspace{2mm}\mathcal{M}=10$}
        \end{center}
      \end{minipage}

      % 2
      \begin{minipage}{0.5\hsize}
        \begin{center}
          \includegraphics[clip, width=7.5cm,bb=0 0 1024 768]{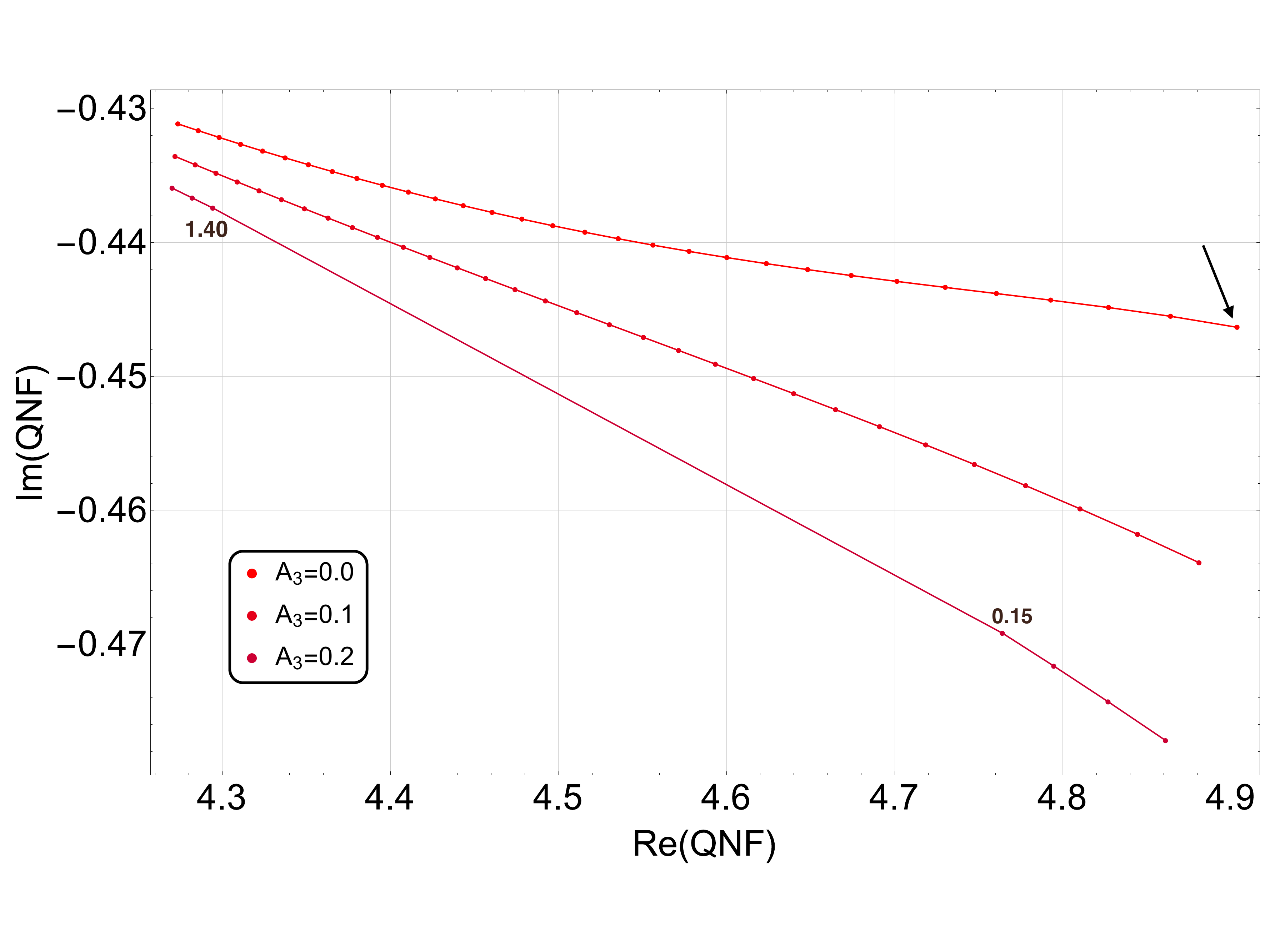}
          \hspace{1.6cm} {\scriptsize $D=8,\hspace{2mm}\mathcal{M}=10$}
        \end{center}
      \end{minipage}
      
      \\
      \\
      \\

      % 3
      \begin{minipage}{0.5\hsize}
        \begin{center}
          \includegraphics[clip, width=7.5cm,bb=0 0 1024 768]{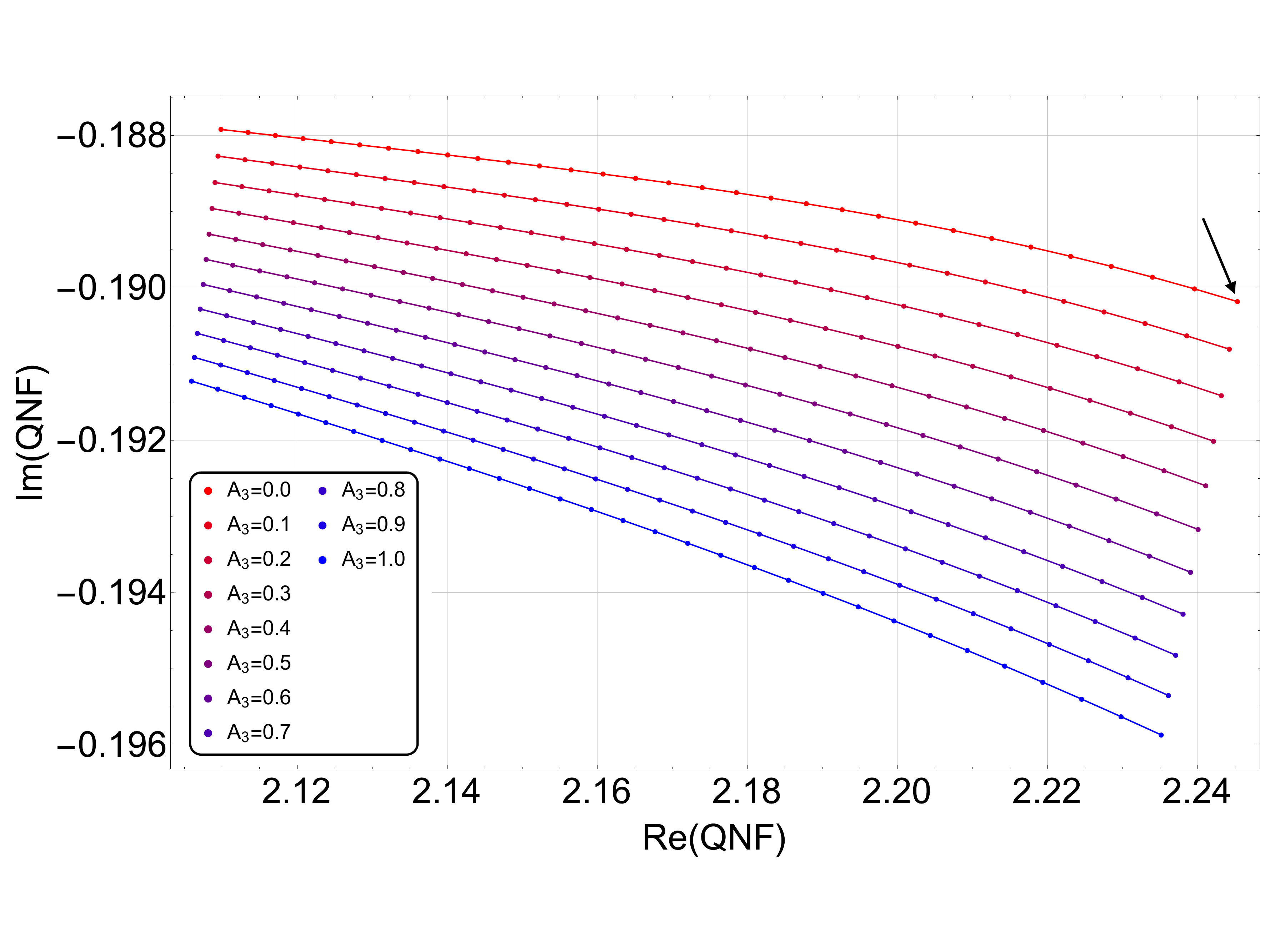}
          \hspace{1.6cm} {\scriptsize $D=7,\hspace{2mm}\mathcal{M}=100$}
        \end{center}
      \end{minipage}

      % 4
      \begin{minipage}{0.5\hsize}
        \begin{center}
          \includegraphics[clip, width=7.5cm,bb=0 0 1024 768]{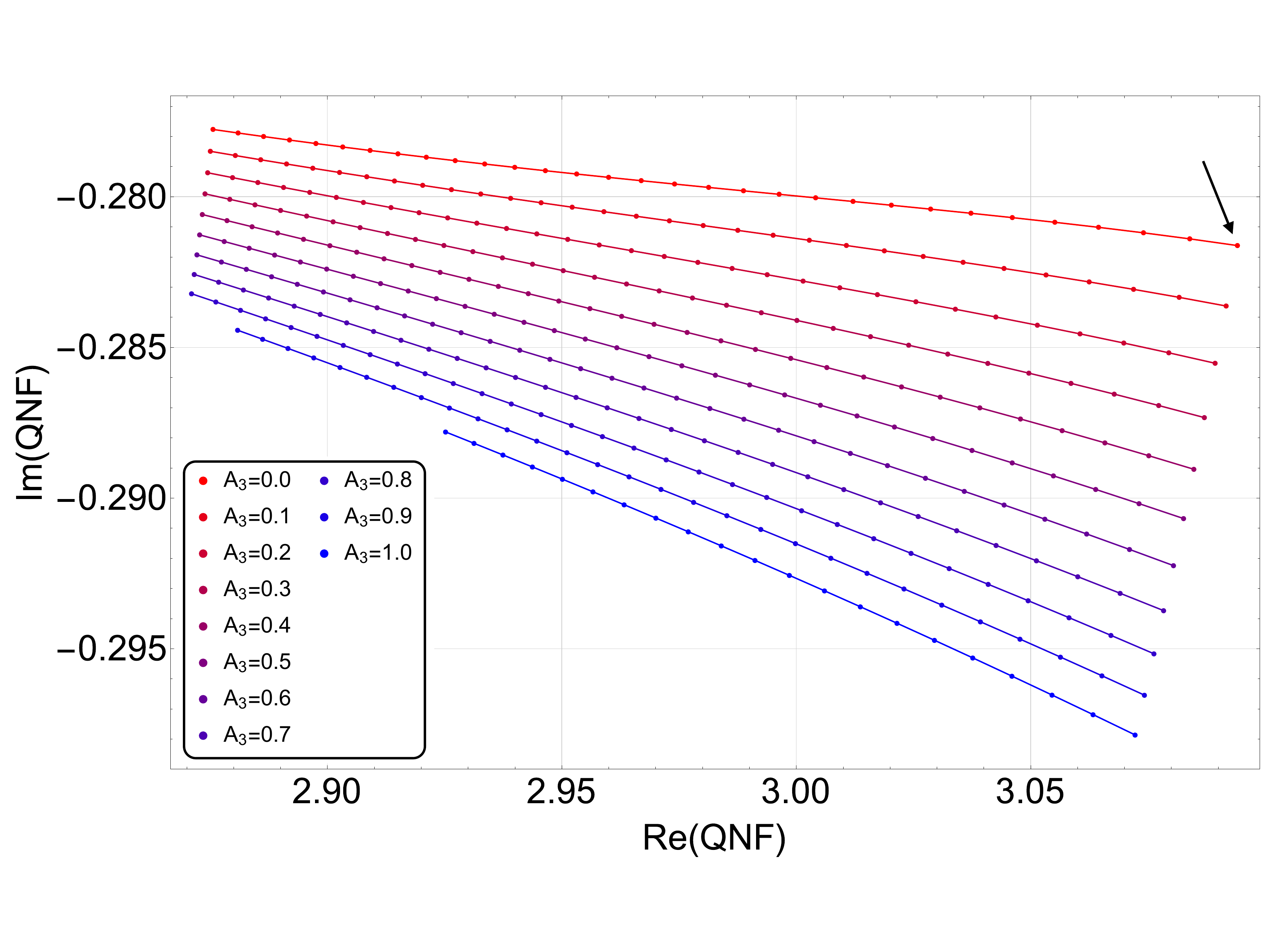}
          \hspace{1.6cm} {\scriptsize $D=8,\hspace{2mm}\mathcal{M}=100$}
        \end{center}
      \end{minipage}
      
      \\
      \\
      \\

      % 5
      \begin{minipage}{0.5\hsize}
        \begin{center}
          \includegraphics[clip, width=7.5cm,bb=0 0 1024 768]{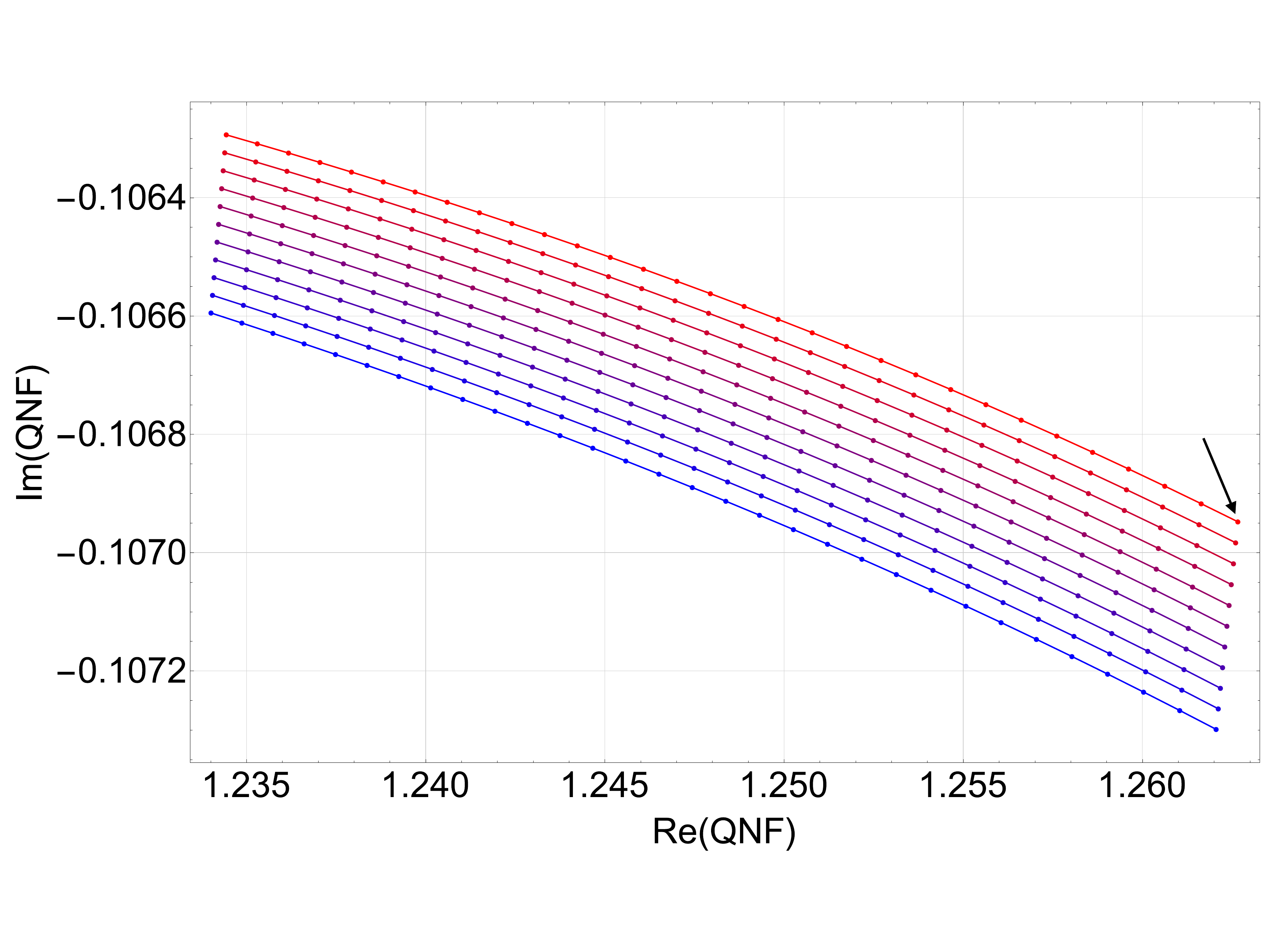}
          \hspace{1.6cm} {\scriptsize $D=7,\hspace{2mm}M=1000$}
        \end{center}
      \end{minipage}

      % 6
      \begin{minipage}{0.5\hsize}
        \begin{center}
          \includegraphics[clip, width=7.5cm,bb=0 0 1024 768]{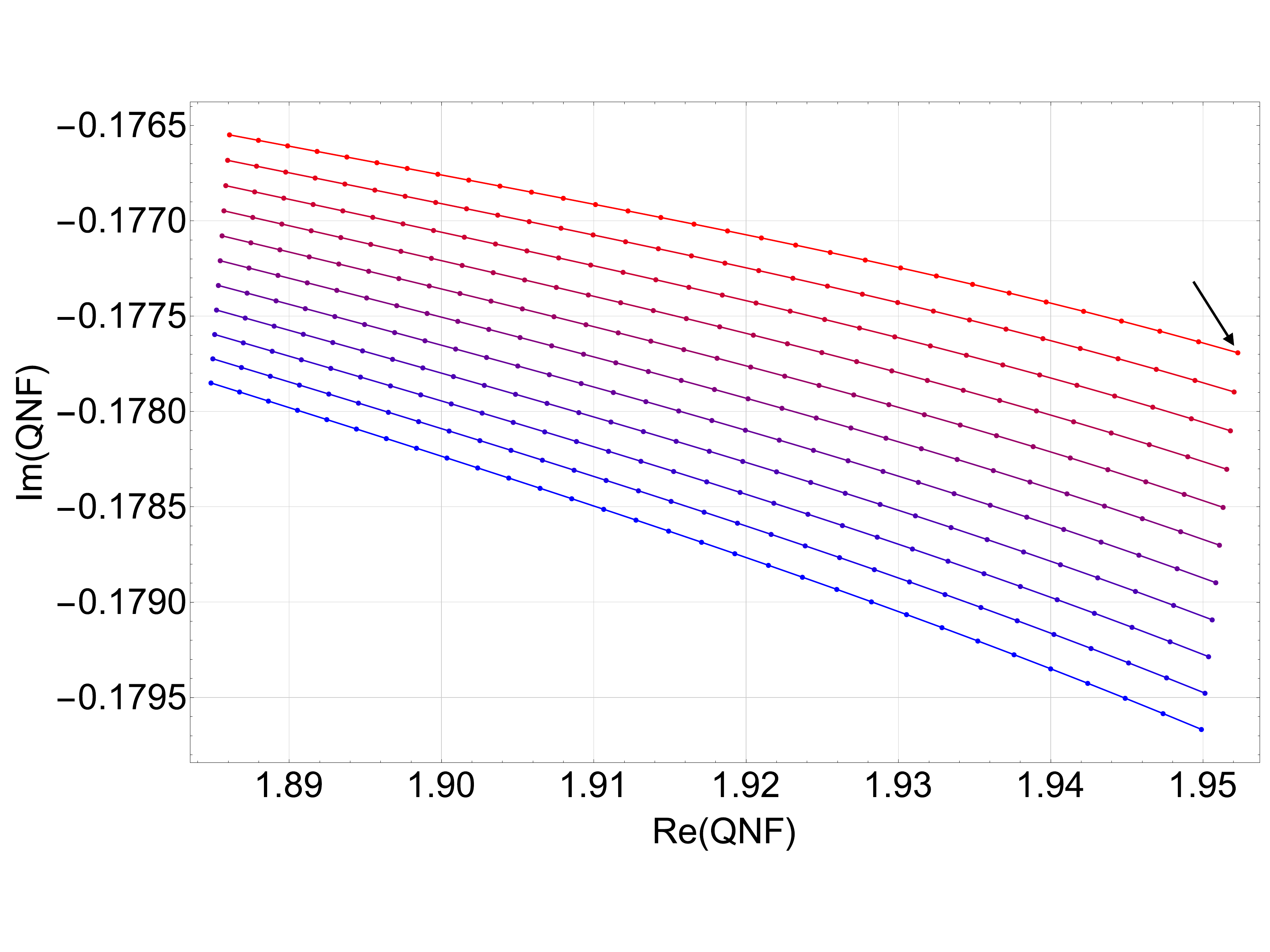}
          \hspace{1.6cm} {\scriptsize $D=8,\hspace{2mm}M=1000$}
        \end{center}
      \end{minipage}

    \end{tabular}

    \label{SP}
  \end{center}
\end{figure}

In Fig.\ref{SF} - Fig.\ref{SP}, we plotted the various QNFs of various fields in Lovelock black holes in seven and eight dimensions.

In Fig.\ref{SF}, we plotted QNFs of the scalar field for the masses ${\cal M}=10, 100,1000$ in seven and eight dimensions.
 In the case of the scalar field, we took the lowest angular parameter $L=0$. 
 In each panel, we plotted a curve by changing the coefficient $A_{2}$ from 0 to 1.5 with the interval of 0.05 for a fixed $A_{3}$.
 Then, we repeated this by changing $A_{3}$ from 0 to 1.0 with the interval 0.1.
 The color of lines in the figures tells us  the value of $A_{3}$.
 The most red color corresponds to the $A_{3}=0$, and the most blue color corresponds to the $A_{3}=1.0$.
 The edge of the curve ( the point indicated by the arrow in the case of the red curve) corresponds to $A_{2}=0$, and the other edge of the same curve corresponds to $A_{2}=1.5$.
 So the QNFs of Einstein gravity in seven and eight dimensions corresponds to the edge of the most red curve indicated by the arrow. 
 In the case of the scalar field, the imaginary part of QNFs seems to converge to a value as we decrease the mass.
 In other words, the Gauss-Bonnet term ceases to be relevant. 
 This trend can be seen clearly in eight dimensions rather than seven dimensions.
 It is also interesting to observe that the pattern is rotating in clockwise.

In Fig.\ref{TP}, we plotted QNFs of the tensor type perturbations of the metric field for the masses ${\cal M}=50, 100,1000$ in seven and eight dimensions.
 We have chosen $L=2$ in the case of tensor type perturbations, which gives the positive definite effective potential with single peak.
 In each panel, as in the case of the scalar field, we plotted a curve by changing the coefficient $A_{2}$ from 0 to 1.5 with the interval of 0.05 for a fixed $A_{3}$.
 Then, we repeated this by changing $A_{3}$ from 0 to 1.0 with the interval 0.1.
 The meaning of the arrow is the same as the case of the scalar field.
 In the case of the tensor type of metric perturbations, we can see there appears a turning point in the curve.
 In this case, initially, the blue curves are always above the red ones irrespective of the mass of black holes.
 As we decrease the mass, the red curve becomes above the blue one for some Gauss-Bonnet parameters.

In Fig.\ref{VP}, we plotted QNFs of the vector type perturbations of metric field for the masses ${\cal M}=10, 100,1000$ in seven and eight dimensions.
 We have chosen $L=10$ in the case of vector type perturbations, which gives the positive definite effective potential with single peak.
 In each panel, as in the case of the scalar field, we plotted a curve by changing the coefficient $A_{2}$ from 0 to 1.5 with the interval of 0.05 for a fixed $A_{3}$.
 Then, we again repeated this by changing $A_{3}$ from 0 to 1.0 with the interval 0.1.
 The meaning of the arrow is  the same as the previous cases.
 In the case of vector type perturbations, there are no special features, which might be related to the fact that there is no instability in this sector. 
 However, in this case, the red curves are always above the blue curves.

In Fig.\ref{SP}, we plotted QNFs of the scalar type perturbations of the metric field for the masses ${\cal M}=10, 100,1000$ in seven and eight dimensions.
 We have chosen $L=10$ in the case of scalar type perturbations, which gives the positive definite effective potential with single peak.
 For the masses ${\cal M}=100,1000$, as in the case of the scalar field, we plotted a curve by changing the coefficient $A_{2}$ with the interval of 0.05 for a fixed $A_{3}$.
 Then, we repeated this by changing $A_{3}$ from 0 to 1.0 with the interval 0.1.
 However, for the masses ${\cal M}=10$, we repeated this by changing $A_{3}$ from 0 to 0.3 with the interval 0.1 in seven dimensions and from 0 to 0.2 with the same interval in eight dimensions.
 The meaning of the arrow is the same as the previous one.
 In the Fig.\ref{SP}, in the case of $D=7,8,M=10$, we have only  sparse plots.
 In the case of the scalar type of metric perturbations as well as the vector type perturbations, the trend is opposite to the tensor type of metric perturbations.
 That is, the Gauss-Bonnet term decreases the absolute value of the imaginary part of QNFs.

Admittedly, the WKB method is not appropriate for a small black hole for which we know there exists the instability.
 We need other methods for these calculations.

\section{Conclusion}
We have studied quasinormal modes of black holes in Lovelock gravity in seven and eight dimensions. 
 Since Lovelock black holes are defined by an algebraic equation, we needed to modify the WKB method.
 Thus, we proposed a novel WKB method  adapted to Lovelock gravity for the calculation of QNFs. 

 As a demonstration, we calculated various QNFs of Lovelock black holes in seven and eight dimensions. 
 We found remarkable features of QNFs of Lovelock gravity which  depend on the coefficients of the Lovelock terms,
 the species of perturbations, and dimensions of the spacetime. 
 In the case of the scalar field, when we increase the coefficients of the third order Lovelock term the real part of QNFs
increase, but the decay rate becomes small irrespective of the mass of the black hole. For small black holes, the decay rate
ceases to depend on the Gauss-Bonnet term.  This tendency  can be seen clearly in eight dimensions than in  seven dimensions.
 The dependence on the Gauss-Bonnet coefficient changes for small black holes. 
 Indeed, there appears a turning point in each curve as the mass becomes small. 
 In the case of tensor type perturbations of the metric field, the tendency of the real part of QNFs is opposite to the scalar field.
 The QNFs of vector type perturbations of the metric field show no particular behavior.
 The behavior of QNFs of the scalar type perturbations of the metric field is similar to the vector type. 
 However, available data are  rather sparse, which indicates that the WKB method is not applicable to many models for this sector.

The features we found in this work are intriguing and worth verifying by using other methods.
 In particular, for calculating QNFs for small black holes, we need to solve partial differential equations numerically
 using the time-domain method. We leave this work for future study.

\begin{acknowledgments}
This work was supported by Grants-in-Aid for Scientific Research (C) No.25400251 and MEXT Grant-in-Aid for Scientific Research on Innovative Areas No.26104708 and MEXT Grant-in-Aid for Scientific Research on Innovative Areas “Cosmic Acceleration” (No.15H05895) . 
\end{acknowledgments}

%\appendix

\end{document}